\documentclass[fleqn,10pt]{wlscirep}
\usepackage[utf8]{inputenc}
\usepackage[T1]{fontenc}
\usepackage{lineno}
\usepackage{tabularx}
\usepackage{multirow}
\usepackage{float} 

\title{A multimodal dataset for understanding the impact of mobile phones on remote online virtual education}

\author[1,*]{Roberto Daza}
\author[2]{Alvaro Becerra}
\author[2]{Ruth Cobos}
\author[1]{Julian Fierrez}
\author[1]{Aythami Morales}
\affil[1]{Biometrics and Data Pattern Analytics Laboratory, Universidad Autonoma de Madrid, School of Engineering, Madrid, 28049, Spain}
\affil[2]{Group for Advanced Interactive Tools, Universidad Autonoma de Madrid, School of Engineering, Madrid, 28049, Spain}

\affil[*]{Corresponding author: Roberto Daza (roberto.daza@uam.es)}

\begin{abstract}

This work presents the IMPROVE dataset, a multimodal resource designed to evaluate the effects of mobile phone usage on learners during online education. It includes behavioral, biometric, physiological, and academic performance data collected from 120 learners divided into three groups with different levels of phone interaction, enabling the analysis of the impact of mobile phone usage and related phenomena such as nomophobia. A setup involving 16 synchronized sensors—including EEG, eye tracking, video cameras, smartwatches, and keystroke dynamics—was used to monitor learner activity during 30-minute sessions involving educational videos, document reading, and multiple-choice tests. Mobile phone usage events, including both controlled interventions and uncontrolled interactions, were labeled by supervisors and refined through a semi-supervised re-labeling process. Technical validation confirmed signal quality, and statistical analyses revealed biometric changes associated with phone usage. The dataset is publicly available for research through GitHub and Science Data Bank, with synchronized recordings from three platforms (edBB, edX, and LOGGE), provided in standard formats (.csv, .mp4, .wav, and .tsv), and accompanied by a detailed guide.

\end{abstract}
\begin{document}

\flushbottom
\maketitle

\thispagestyle{empty}


\section*{Background \& Summary}


Education is a fundamental pillar of society and has continually evolved throughout history. In recent years, it has been deeply influenced by the digital revolution, which has moved our society from the physical to the digital world. Evidence of this change in recent decades includes the growing significance of e-learning or online education, expected to grow exponentially over the next 20 years \cite{daza2023edbb, GMI_eLearning2024}. Relevant institutions are adopting this model, especially those offering MOOCs (Massive Open Online Courses), due to their ability to reach a broader learner base \cite{ma2019investigating}. 

Learners have also been significantly impacted by the digital revolution, which has altered their habits to incorporate electronic devices into their daily routines such as mobile phones and tablets. Since 2012, studies \cite{tindell2012use, huey2023impact} have indicated  that between 97\% and 99\% of learners own a mobile phone. Moreover, 95\% of learners bring their phones to class every day, 92\% use these devices to send text messages during class, and 10\% even admit to texting during exams. On average, learners spend about five hours per day on their mobile phones, predominantly using messaging services (sending an average of 109 text messages per day) or social networks like WhatsApp and Instagram \cite{andrews2015beyond,smith2011americans}.

Therefore, the mobile phone raises numerous questions: Can it become a distraction to learners’ attention? Or, conversely, could the absence of a mobile phone cause greater distraction or anxiety? Considering that attention is defined as the ability to concentrate, specifically, to exert conscious cognitive effort on a particular task or stimulus at any given moment \cite{daza2024deepface}, and given that various studies suggest it is easy to lose focus due to short-term distractions such as messages, sounds, etc.~\cite{altmann2014momentary, tanil2020mobile}, it is reasonable to consider how mobile phones could impact learner performance.  This is not a new idea, and various studies have shown that mobile phone use negatively affects learners’ attention, memory, and academic performance. This is mainly because it increases cognitive load by inducing learners to multitask~\cite{tanil2020mobile, huey2023impact, mendoza2018effect, froese2012effects}. This effect becomes more noticeable as the class progresses or during long-duration activities~\cite{mendoza2018effect}. However, learners’ performance can be influenced by their emotional state. For this reason, nomophobia, defined as the fear, stress, or anxiety caused by being unable to access or use a mobile phone, is an important factor to consider. Various studies have demonstrated how this condition affects memory and attention, ultimately impacting learners’ academic performance~\cite{cheever2014out,mendoza2018effect}.  

Interestingly, although nomophobia can lead to reduced academic performance, the previously mentioned studies have shown that just the presence of a mobile phone also diminishes learners’ attention levels, presenting a complex challenge that requires in-depth investigation.

The IMPROVE dataset was collected in a realistic online learning environment. A total of 120 learners were monitored, divided into three distinct groups to understand the effects of mobile phone usage. To our knowledge, IMPROVE dataset is the first to provide a comprehensive multimodal acquisition setup including behavioral \cite{daza2023edbb,hernandez2019edbb}, biometric, and learning analytics data from learners during an authentic online course that considers mobile phone usage. The dataset aims to facilitate understanding of the effects of mobile phone usage in online education. Additionally, it provides a robust dataset that can be utilized to develop new technologies to enhance learners’ learning processes or ensure secure certification \cite{hernandez2019edbb, baro2018integration, bhattacharjee2018enhancing, cobos2023self, topali2024codesign, daza2025smartevr}.

\begin{figure*}[t]
    \centering
    \includegraphics[width=\textwidth]{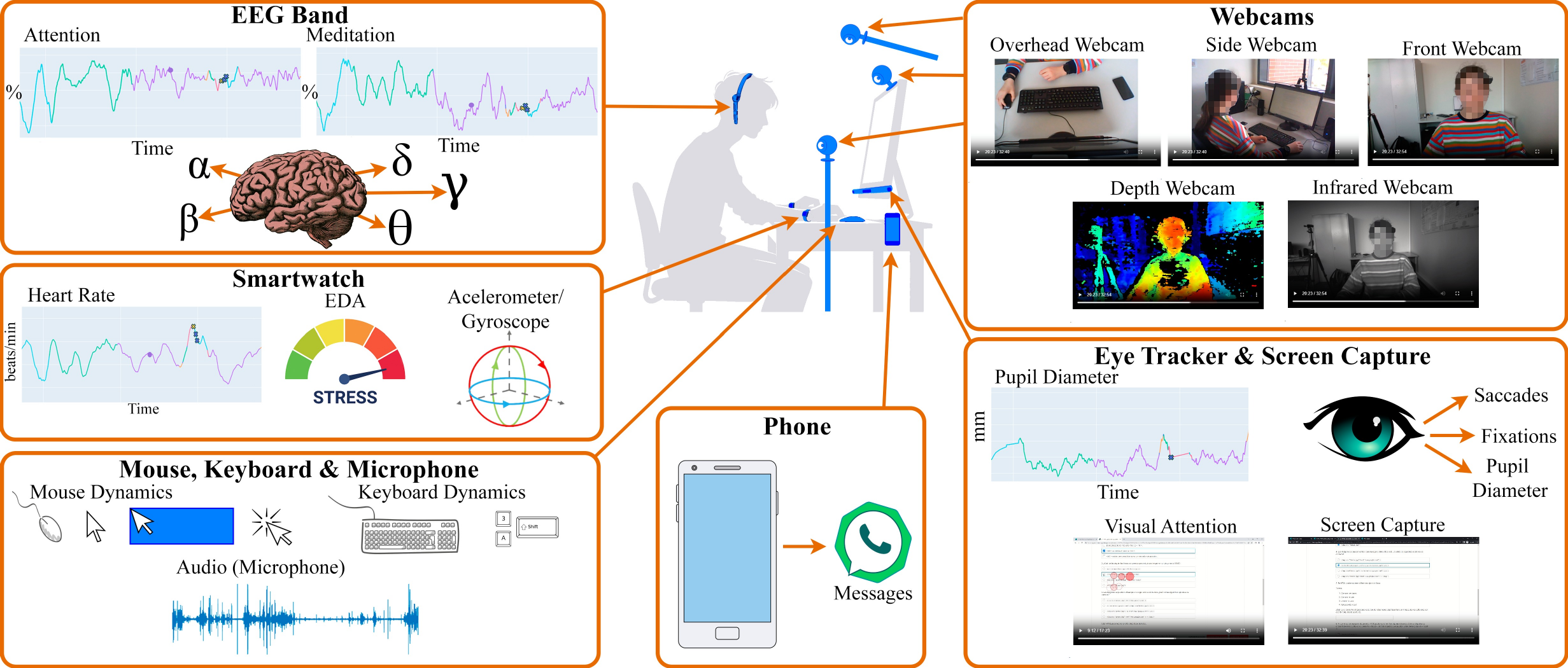} 
    \caption{Acquisition setup used during data capture with the edBB\cite{daza2023edbb, hernandez2019edbb} platform, illustrating all sensors and devices employed, along with examples of the captured information.}
    \label{fig:Setup}
\end{figure*}

We employed 16 sensors, recognized as effective indicators for understanding learner behavior \cite{nagao2023virtual, daza2024deepface, zhou2024stat, daza2024mebal2}, including an electrodermal activity (EDA) sensor, an electroencephalogram (EEG) band, two pulse meters, an accelerometer, a gyroscope, three RGB cameras, two Near-Infrared (NIR) cameras, a mouse, a keyboard, a microphone, a monitor, and an eye tracker (see Fig.~\ref{fig:Setup}).
These sensors enabled us to monitor cognitive activity, biometric, and behavioral signals throughout 120  online learning sessions, each lasting approximately 30 minutes. During these sessions, learners interacted with educational content related to HTML, including video lectures, document reading, and test completion to assess their knowledge both before and after the session (with a total of 16 questions per session). This comprehensive setup allowed us to track not only how learners engaged with the material but also how external factors, such as mobile phone interruptions, affected their focus and learning outcomes.

In total, the dataset comprises 2.83 terabytes of data, making it one of the most extensive multimodal datasets in this domain. It includes  145 labeled mobile phone interruptions, providing crucial insights into how mobile phone usage affects learners.

\section*{Methods}



This section outlines the experimental methods and materials used for data collection. The dataset was obtained in accordance with the Declaration of Helsinki and was approved by the Ethics Committee of the Universidad Autónoma de Madrid (Approval No. CEI-130-2699, granted on 11/04/2023). Learners were informed about the experimental procedure and the data acquisition setup, and each participant signed a written consent form.   The form clearly stated that the collected data would be used exclusively for research purposes and included in a publicly available dataset intended for the research community.  Furthermore, participants gave explicit consent for the publication of non-anonymised videos showing their faces. Learners received an incentive for participating.

\subsection*{Dataset: IMPROVE} \label{sec:Database}




The dataset consists of 120 learners from the School of Engineering at the Universidad Autónoma de Madrid (UAM), who were monitored during a real online learning session focused on HTML, lasting between 20 and 35 minutes. The session was part of the  MOOC  titled “Introduction to Development
of Web Applications” (WebApp MOOC for short) available on the edX platform (\url{https://www.edx.org}).

\begin{table}[h!]
\centering
\begin{tabular}{|l|l|c|c|c|}
\hline
\textbf{Category} & \textbf{Subcategory} & \textbf{Percentage} & \textbf{Number of Learners} & \textbf{Average Age} \\ \hline \hline
\multirow{5}{*}{Degree Program} 
& Computer Engineering & 32.5\% & 39 & 21.77 \\ \cline{2-5}
& Telecommunications Engineering & 30.0\% & 36 & 22.06 \\ \cline{2-5}
& Biomedical Engineering & 25.0\% & 30 & 19.80 \\ \cline{2-5}
& Data Science and Engineering & 3.3\% & 4 & 20.00 \\ \cline{2-5}
& Computer Engineering and Mathematics & 9.2\% & 11 & 20.55 \\ \hline \hline
\multirow{4}{*}{HTML Proficiency} 
& None & 42.5\% & 51 & 20.02 \\ \cline{2-5}
& Beginner & 35.8\% & 43 & 21.95 \\ \cline{2-5}
& Intermediate & 16.7\% & 20 & 21.80 \\ \cline{2-5}
& Advanced & 5\% & 6 & 23.67 \\ \hline \hline
\multirow{5}{*}{Medical Conditions} 
& No issues &  --- & 55 & --- \\ \cline{2-5}
& Glasses &  --- & 41 & --- \\ \cline{2-5}
& Contact lenses &  --- & 19 & --- \\ \cline{2-5}
& Heart murmur &  --- & 2 & --- \\ \cline{2-5}
& Myopia without correction &  --- & 4 & --- \\ \hline \hline
\multirow{3}{*}{Overall Averages} 
& Overall Average Age & --- & --- & 21.19 \\ \cline{2-5}
& Average Age of Male Learners & 51.83\% & 61 & 21.64 \\ \cline{2-5}
& Average Age of Female Learners & 49.16\% & 59 & 20.73 \\ \hline
\end{tabular}
\caption{Distribution of learners by degree program, HTML proficiency levels, medical conditions (percentages are not shown because learners can have more than one condition), and gender, along with their average ages.}
\label{table: Distribution} 
\end{table}

\subsubsection*{Recruitment.}

Learners from the School of Engineering at the Universidad Autónoma de Madrid (UAM) were recruited. The School of Engineering was selected because its five degree programs all require basic knowledge of computer science, which aligns well with the online course. An economic incentive was offered to encourage participation. A total of 170 learners signed up for the study, of whom 120 were selected. An initial pilot test was conducted with six learners to verify the experimental protocol and monitoring system.  Adjustments were made after the initial test, including: requiring Group 1 learners to provide more detailed responses to messages (for more information about the groups formed, see the subsection Protocol), giving learners a more comprehensive explanation of the session beforehand, removing an initial form to avoid potential suggestion effects, and making technical adjustments to the EEG band and the eye tracker calibration. The six learners from the pilot test did not participate in the final data collection. Four additional learners were excluded due to issues during the session and replaced by other participants. The final 120 learners were divided into three groups of 40. The selection process considered several criteria to ensure balanced groups, including gender (61 male, 59 female), HTML proficiency (none, beginner, intermediate, advanced), degree program, among other factors. Learners ranged in age from 18 to 32 years, with an average age of 21 (see Table~\ref{table: Distribution}). All learners gave their informed consent for the publication of the dataset. Each learner was anonymized with an identifier that cannot be linked to their identity.

\subsubsection*{Protocol.} The objective was to examine the effects of mobile phone presence or absence on learner behavior and academic performance. To this end, three distinct groups were formed:
\begin{enumerate}
    \item \textbf{Group 1:} Mobile phone use and possession were allowed. The device was placed on the learner’s desk, visible to the learner, with both sound and vibration activated.
    \item \textbf{Group 2:} Mobile phone possession was allowed, while use was prohibited. The device was placed on the learner’s desk with the screen facing down, and sound and vibration were activated.
    
    \item \textbf{Group 3:} The mobile phone was confiscated for the entire duration of the learning session.
\end{enumerate}

\begin{figure*}[t!]
    \centering
    \includegraphics[width=0.95\textwidth]{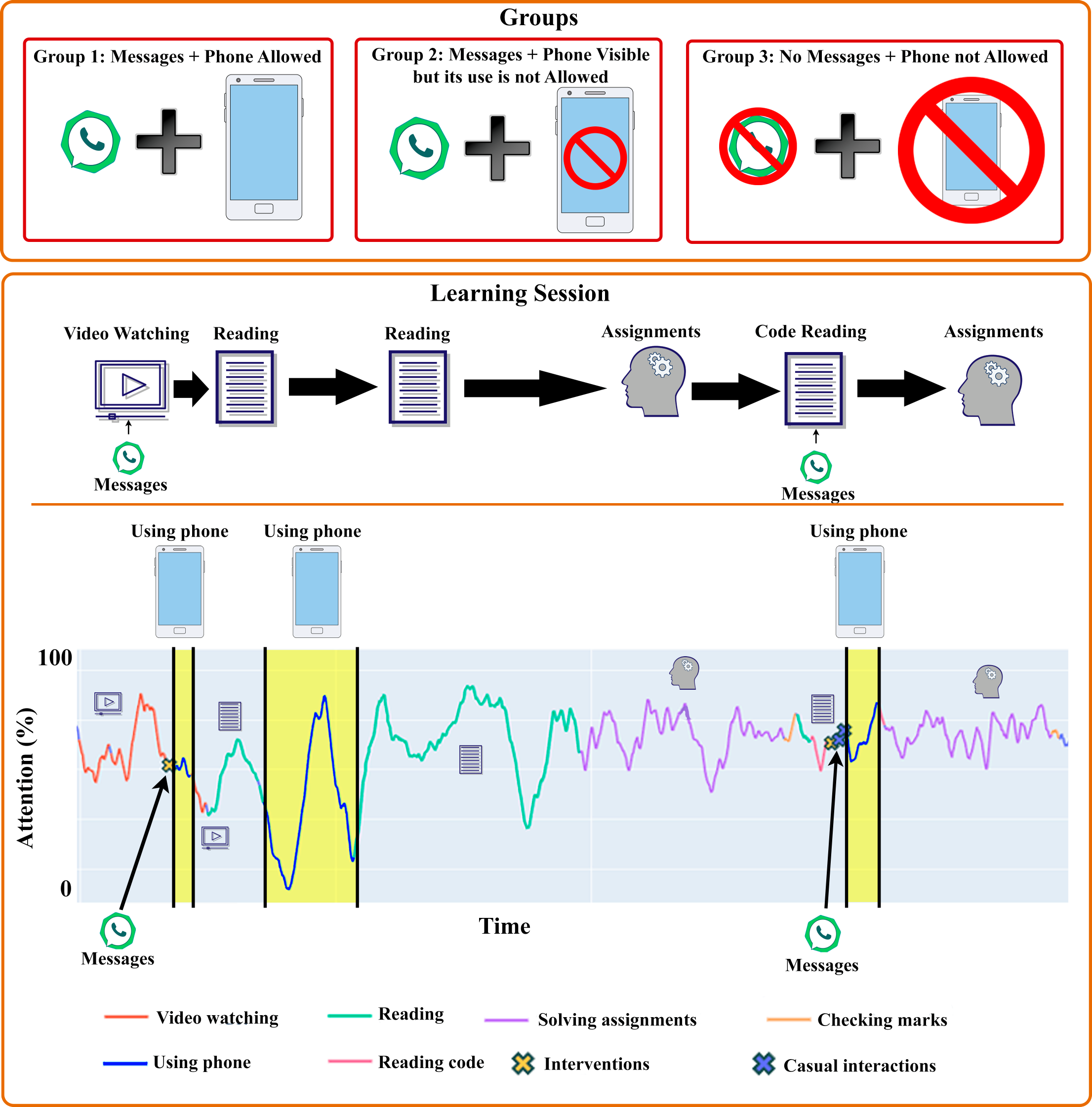} 
    \caption{Protocol followed during the learning session to evaluate the impact of mobile phone usage on learners’ academic performance. The top section presents the experimental design, illustrating the division of learners into three groups based on the permitted level of mobile phone interaction: unrestricted use, visible but prohibited use, and complete removal. The bottom section details the structure of the learning session, including the sequence of activities (e.g., video watching, reading, and completing assignments). It also shows a sample graph of a learner’s attention signal, synchronized with the session timeline. Time intervals in the signal are color-coded to reflect each activity, each accompanied by its corresponding icon, while windows of phone usage are highlighted by vertical boundary lines, with contextual icons indicating the type and timing of interruptions. This provides a visual representation of when phone-related interruptions occurred during the session.
    }
    \label{fig:LS}
\end{figure*}

To ensure consistency across all groups, all learners underwent the same learning session, delivered in a fixed sequence without enforced time limits between tasks. Learners were otherwise free to behave as they typically would in an e-learning context, with the only variation stemming from the mobile phone usage condition assigned to each group. No explicit instructions were given to avoid distractions unrelated to mobile phone use. However, two general restrictions applied: (i) learners could not return to previous tasks, and (ii) they were not allowed to control video playback (i.e., pause, rewind, or skip).

Three supervisors were present throughout the session. The first supervisor was responsible for explaining the session in detail to the learner, describing the operation of the sensors, placing and calibrating them, and ensuring that the platforms functioned correctly. The second supervisor used the LOGGE tool \cite{becerra2023m2lads} (for more details on this tool, see the "Data, platforms and sensors" subsection) to label the moments when the learner interacted with the mobile phone or when it rang, as well as to label other relevant information for the session. The third supervisor was responsible for sending two messages to the learners at key moments during the learning session for both Group 1 and Group 2, reviewing the information recorded by the second supervisor, noting any possible incidents during the session, and assisting the learner in case of doubts. We will differentiate between two different types of interruptions: interventions (the two controlled messages sent by the supervisor) and casual interactions (uncontrolled messages/calls received by the learner that were not related to the interventions). Therefore, interventions are considered controlled events, while interactions are classified as uncontrolled. The interventions were always delivered at the same points during the session. The first intervention occurred during an explanatory video of the lesson, and the second during the reading of HTML code, which was part of the learner evaluation (see Fig.~\ref{fig:LS}). Learners in Group 1 were required to respond with a detailed reply. The messages consisted of simple questions such as “What do you think of the cafeteria of our School of Engineering?”, or an image from a popular movie like Harry Potter, accompanied by the question: “What do you think of this film series? Would you recommend it?” On average, learners took about 26 seconds to respond to the first message and about 36 seconds to respond to the second message. Learners in Group 2 were not allowed to use their mobile phones, so their phones simply rang and vibrated during the interruptions.

Additionally, learners in Group 1 were given the freedom to respond/send messages or calls to their contacts. Each of these interactions was recorded. 
On average, the learners in Group 1 chose to interact with their mobile phones approximately 5 times (including controlled messages and uncontrolled messages or calls), while the mobile phone in Group 2 rang/vibrated approximately 7 times.

\subsubsection*{Tasks}

Each learner completed a single learning session focused on HTML, taken from the WebApp MOOC. The session lasted an average of 30 minutes,  with most learners completing it in 20 to 35 minutes, and a maximum duration of 40 minutes. The tasks performed during the session were as follows:

\paragraph{Before starting the HTML unit:}
\begin{itemize}
    \item \textbf{Calibration and Learner Information:} Learners were briefed again on the session procedures. Sensors, such as the eye tracker and EEG band, were fitted and calibrated. Learners were asked to provide information that could affect the sensors, such as the use of glasses/contact lenses, cardiac conditions, handedness (right or left-handed), etc. Specific medical conditions reported by the learners, including their prevalence, are summarized in Table \ref{table: Distribution}.
    
    \item \textbf{EDA Test:} Before starting the session, an initial EDA  test was conducted to understand the learners’ stress levels. Learners were also asked to self-report their current state as "Very stressed, stressed, normal, relaxed, extremely relaxed" to obtain their feedback.
    
    \item \textbf{Form and Pretest:} Additionally, they  filled out a form with the following two questions, i) What is your current level of stress/anxiety? and ii) What is your current level of distraction? This information was collected to gather learner feedback prior to the session and to validate it against sensor data. Learners also completed a pretest consisting of eight multiple-choice questions of low to intermediate difficulty on HTML, with two objectives: \textit{i)} to assess the learners’ current level of HTML knowledge and \textit{ii)} to balance the three groups. 
\end{itemize}

\paragraph{HTML Unit:}
The HTML unit consisted of a theoretical part where learners were introduced to HTML concepts in web design through explanatory videos and reading web documents, and a second part focused on assessing their understanding. This evaluation involved eight multiple-choice questions. After completing the test, learners were able to review their incorrect answers. At no point were learners allowed to return to previous activities, pause the videos, or skip forward/backward in the video. The tasks were ordered as follows (see Fig.~\ref{fig:LS}):
\begin{enumerate}
    \item \textbf{Videos:}  Two introductory videos on interactive web and basic HTML concepts. The first video lasted 20 seconds, while the second lasted 1 minute and 48 seconds. During the second video, a message was sent to the learners in Groups 1 and 2 after 1 minute, coinciding with the explanation of HTML concepts.
    \item \textbf{Document Reading:} Two documents further explaining the videos, covering concepts such as HTML structure, different tags, etc.
    \item \textbf{Test:} This test was designed to be compared with the pretest, in order to evaluate whether the students had learned during the learning session compared to their initial knowledge. The questions were similar to those in the pretest. Six multiple-choice questions followed by a phase of reading HTML code to answer two more multiple-choice questions. During the HTML code reading, the second message was sent to learners at the beginning of the reading when comprehension of the code required greater attention. However,  learners could resume reading the code after answering the message, unlike the video where they could not rewind.

    \item \textbf{Review Phase:} Finally, learners accessed a review phase where they could check and understand the mistakes they made.
\end{enumerate}

\begin{table}[htbp]
    \centering
    \begin{tabular}{|c|c|c|c|c|c|c|c|c|c|}
    \hline
    \multirow{2}{*}{\textbf{Learner Feedback}} & \multicolumn{3}{c|}{\textbf{Group 1}} & \multicolumn{3}{c|}{\textbf{Group 2}} & \multicolumn{3}{c|}{\textbf{Group 3}} \\ \cline{2-10} 
    & \textbf{Low} & \textbf{Medium} & \textbf{High} & \textbf{Low} & \textbf{Medium} & \textbf{High} & \textbf{Low} & \textbf{Medium} & \textbf{High} \\ \hline
    Anxiety at Start &52.63 & 39.47 & 7.89 & 51.43 & 40.0 & 8.57 & 43.24 & 51.35 & 5.41 \\ \hline
    Distraction at Start & 47.37 & 52.63 & 0.0 & 51.43 & 34.29 & 14.29 & 67.57 & 21.62 & 10.81 \\ \hline
    Anxiety During Session & 60.53 & 28.95 & 10.53 & 60.0 & 34.29 & 5.71 & 64.86 & 35.14 & 0.0 \\ \hline
    Distraction During Session & 44.74 & 39.47 & 15.79 & 42.86 & 45.71 & 11.43 & 72.97 & 21.62 & 5.41 \\ \hline
    Session Difficulty & 37.5 & 55.00 & 7.5 & 40.0 & 42.5 & 17.5 & 45.0 & 47.5 & 7.5 \\ \hline
    Performance Level & 7.89 & 50.0 & 42.11 & 8.57 & 60.0 & 31.43 & 10.81 & 43.24 & 45.95 \\ \hline
    Phone Distraction & 26.32 & 50.0 & 23.68 & 40.0 & 50.0 & 10.0 & - & - & - \\ \hline
     Stress Due to Phone Inaccessibility & - & - & - & 73.68 & 26.32 & 0.0 &  95.0 & 5.0 & 0.0 \\ \hline
    
    \end{tabular}
    \caption{Learner feedback from forms, categorized into three levels (low, medium, and high) and expressed as percentages. The responses cover aspects such as anxiety, distraction, perceived difficulty, and self-assessed performance, and are presented separately for each experimental group.}
    \label{tb:Forms}
\end{table}

\paragraph{End of the Session:}
\begin{itemize}
    \item A final \textbf{EDA Test} was conducted to detect post-session stress levels, and learners’ feedback was collected again.
     \item Learners completed a post-session \textbf{Form}, which asked about their levels of stress, distraction, perceived difficulty, and performance during the session. Additionally, learners in Groups 1 and 2 were asked about phone distraction, while Groups 2 and 3 were asked about stress from phone deprivation. Table \ref{tb:Forms} presents the responses collected from the self-reported forms.

    \item Learners were asked to fill out a form with the following six questions:
    \begin{itemize}
        \item "What was your level of stress/anxiety during the learning session?"
        \item "What was your level of distraction during the learning session?"
        \item "How would you evaluate the difficulty of the learning session you just completed?"
        \item "How would you rate your performance during the learning session?"
        \item "Observations: Please add any comments you wish to make after the data collection."
    \end{itemize}
    
    The final question varied depending on the group assigned to each learner:
    \begin{itemize}
        \item For Groups 1 and 2: "How distracted were you by your mobile phone?"
        \item For Groups 2 and 3: "How stressed did you feel due to not being allowed to use your mobile phone?"
        
    \end{itemize}
\end{itemize}




\begin{table}[h!]
\centering

\begin{tabularx}{\textwidth}{|>{\hsize=0.9\hsize}X|>{\hsize=0.9\hsize}X|>{\hsize=0.6\hsize}X|>{\hsize=1.6\hsize}X|}
\hline
\textbf{Information Type} & \textbf{Sensors or Platforms} & \textbf{Sampling Rate} & \textbf{Features} \\ \hline \hline
Video & 3 RGB cameras \newline 2 Infrared cameras \newline 1 Depth camera & 20 Hz - 30 Hz & MP4 files with codec H264 \\ \hline
Desktop Video & Screen & 1 Hz & MP4 file with codec H264 \\ \hline
Audio & Microphone & 8000 Hz & Uncompressed WAV files \\ \hline
Keystroke & Keyboard & 12 Hz & Keypress and key release events \\ \hline
Mouse Dynamics & Mouse & 895 Hz & Mouse events: Move, press/release, drag and drop and mouse wheel spin \\ \hline
EEG & Band & 1 Hz & Power Spectrum Density of five frequency bands. Level of attention (from 0 to 100) and eyeblink strength \\ \hline
Pulse, Stress, temperature and Inertial & 2 SmartWatch: Huawei Watch 2 and FitBit Sense with PPG, Gyroscope,  Accelerometer, Electrodermal Activity & 100 Hz & Timestamps and data from the pulse, EDA sensor (skin conductance), body temperature and the inertial sensors (accelerometer, gyroscope, magnetometer) \\ \hline

 Visual attention, Gaze, and Eyeblink & 2 Eye Tracking Cameras & 120 Hz & Gaze origin and point, saccades, fixations, gaze event duration, pupil diameter, eyeblink, data quality

\\ \hline

Face position, Head Pose and Facial Landmarks  & edBB platform & 30Hz & Facial bounding box information,  468 facial landmarks, Euler angles
 
\\ \hline
Context Data & edBB platform, LOGGE Data-M2LADS, edX,  learner, Forms & NA & Learner Information (gender, HTML proficiency level, field of study, health issues, etc.), mobile phone usage events, computer details including name, private IP, public IP, MAC address, operating system (OS), architecture, keyboard language, screen resolution, available memory, total memory, start and finish times of each task session, and test answers, etc.

\\ \hline
\end{tabularx}
\caption{ \label{table: Sensors} Summary of sensors and data collection from the IMPROVE dataset.}
\end{table}

\setlength{\extrarowheight}{2pt} 

\subsubsection*{Data, Platforms and Sensors}\label{sec:data_platforms_sensors}
Two different platforms were used to monitor the learners: the edX and edBB platforms \cite{daza2023edbb,hernandez2019edbb}.
The edX platform is an online MOOC platform used to conduct the learning sessions. It also allows for the capture of session metadata from learners, such as test answers, task start and end times, and more.
The edBB platform was used to monitor behavioral and biometric signals that have been shown to contain relevant information for understanding learner engagement and performance \cite{daza2023edbb,hernandez2019edbb, daza2024mebal2, daza2024deepface, nagao2023virtual, daza2021alebk, rafiqi2015pupilware }. See Daza et al.~\cite{daza2023edbb} for a video demonstration (\url{https://www.youtube.com/watch?v=JbcL2N4YcDM}). A multimodal acquisition framework was designed, incorporating 16 sources of information/sensors (see Fig.~\ref{fig:Setup} and Table~\ref{table: Sensors}):
\begin{itemize}
\item Cameras: Two Logitech C170 cameras (side and overhead) operating at 20 Hz with a resolution of $640 \times 480$, and one front-facing Intel RealSense camera were used. The RealSense\cite{intel2020realsense} camera includes one RGB camera and two NIR cameras, with dimensions of 90 mm length $\times$ 25 mm (depth) $\times$ 25 mm (height). The NIR cameras are monochrome and sensitive to both the visible spectrum and NIR, following the sensitivity curve of the CMOS sensors. The three Intel RealSense cameras were configured to operate at 30 Hz and a resolution of $1280\times720$, and depth images were obtained by combining their three image channels. See Fig.~\ref{fig:Setup} for more details on the position of each camera.

\item Smartwatches: Two different smartwatches, the Huawei Watch 2 and the FitBit Sense, were used to monitor heart rate \cite{hosseini2022multimodal, hernandez2020heart, hernandez2019edbb}, stress levels via the electrodermal activity (EDA) sensor \cite{hosseini2022multimodal, cocskun2023physiological, romero2023ai4fooddb}, and inertial data using gyroscopes and accelerometers \cite{acien2022becaptcha}. Two different smartwatch models were used, each worn on a different wrist, so that the heart rate monitored by the devices could be compared. The heart rate measurement was set to 1 Hz for the Huawei Watch 2 and 0.2 Hz for the FitBit Sense. The FitBit Sense has one sensors that are not available in the Huawei Watch 2: the EDA sensor. The Huawei Watch 2 was worn on the dominant hand of the learner to detect the movements produced by mouse usage.


\item EEG: A NeuroSky EEG headset, was used to measure the power spectral density across five electroencephalographic frequency bands ($\alpha, \beta, \gamma, \delta, \theta$). Through the preprocessing of these EEG channels, estimates of attention and meditation levels, as well as the occurrence of eyeblinks, are obtained. EEG captures voltage signals typically generated by synaptic excitations of the dendrites in the pyramidal cells located in the top layer of the brain cortex \cite{kirschstein2009source}. These signals are primarily produced by the synchronous firing of numerous neurons and fibers \cite{hall2020guyton}. 
EEG is an important, effective, and objective measure for estimating cognitive load and learner states \cite{chen2018effects, li2011real} because these signals are sensitive to the mental effort/cognitive works and mental states such as learning,  lying, perception and stress.
Extensive research in neurology and learning fields has explored the relationship between EEG signals and mental activities \cite{daza2021alebk, daza2023matt, hall2020guyton, lin2018mental, chen2017assessing, li2009towards}, demonstrating the potential of the five electroencephalographic channels and the correlation between eyeblinks and cognitive load \cite{daza2020mebal, daza2024mebal2, daza2021alebk, bagley1979effect}. In the context of this dataset, EEG data were included to provide a direct and continuous measurement of learners’ cognitive states throughout the session. This information constitutes a key component for evaluating the impact of mobile phone usage on cognitive load, particularly with regard to fluctuations in attention levels and potential effects associated with nomophobia or mobile-induced distractions. This is especially relevant given that attention is defined as the cognitive effort directed toward a task and plays a pivotal role in ensuring accurate comprehension during learning.

\item Visual attention: A Tobii Pro Fusion eye tracker equipped with two high-speed infrared cameras configured at 120 Hz for eye tracking. This device estimated the following data: gaze origin and point, pupil diameter, eye movement type (fixation, saccade, unclassified, eyes not found), event duration, data quality, eyeblink, and more; allowing us to measure visual attention.
Previous studies have demonstrated that visual attention is a key factor in learning, providing a deeper understanding of learner performance \cite{sharma2016gaze, sharma2016visual}. There is a strong correlation between learner performance and visual attention, with features such as fixations, pupil diameter, eyeblink and saccades being closely associated with learning outcomes \cite{sharma2016gaze}, fatigue detection \cite{andreu2016ealab}, and task prediction \cite{Navarro2024}.

\begin{figure*}[t]
    \centering
    \includegraphics[width=\textwidth]{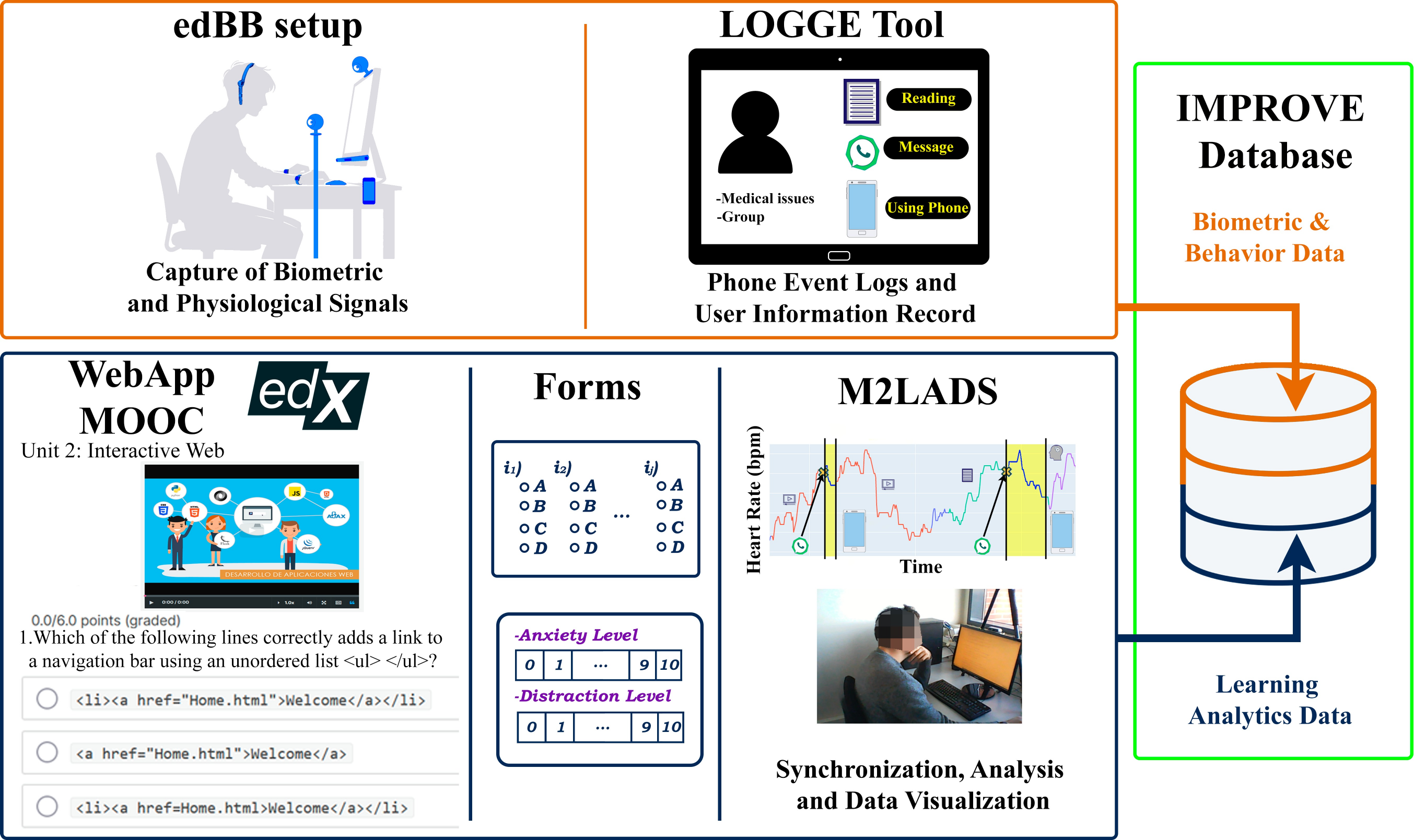} 
    \caption{Overview of the platforms and tools used to capture the IMPROVE dataset. The edBB setup was used to monitor the learner’s biometric and behavioral signals during online sessions. The edX platform delivered the course content and collected session metadata. The LOGGE Tool recorded additional information, such as medical issues and group assignments, and mobile phone usage. The data were aggregated into the IMPROVE dataset, which could be analyzed using the M2LADS platform to visualize and synchronize the captured signals and events.}
    \label{fig:Platforms}
\end{figure*}

\item Computer Information: Relevant information for the session were obtained from various sources on the learner’s computer.
\textit{i)} Microphone Information: Audio was recorded from the computer’s microphone at a  sampling rate of 8000 Hz. The audio was recorded throughout the session, and all learners pronounce an initial phrase, "Sound test for the edBB platform," which can be used as a baseline for future research. \textit{ii)} Keyboard and Mouse Activity: Keystrokes, mouse position,  inter-keystroke timing, mouse wheel movements, etc., are monitored. Keystrokes and mouse dynamics have proven to be effective mechanisms for understanding learner sessions and providing secure certification \cite{morales2016kboc,2016_IEEEAccess_KBOC_Aythami}. \textit{iii)} Screen Capture: The monitor screen was recorded at a frequency of 1 Hz.
iv) Session metadata: Information such as keyboard type, logging data, IP and MAC addresses, operating system, etc., was collected.
\end{itemize}
The edBB platform facilitated optimal synchronization across all sensors, recording the captured samples in both UNIX time and local time, and sending the information to a common server for easy access. By using a single platform to control all sensors, it was possible to synchronize the activation and termination times of each sensor. This configuration also enabled the acquisition of precise information, such as the exact moment a learner pressed a key and the corresponding video frame, without requiring additional post-processing.

In addition to using these platforms, a logging tool called LOGGE \cite{becerra2023m2lads} was developed with the primary goal of labeling the mobile phone events for Group 1 and Group 2, which consisted of learner interactions with their mobile phones or when their phones rang. Additionally, this tool collected learner-related information (e.g., use of glasses/contact lenses, cardiac conditions, handedness, etc.) and enhanced the edX log traces with details about the activities performed during the learning sessions, such as task start and end times, error logs, and more.

Finally, a web-based system called M2LADS \cite{becerra2023m2lads, BecerraM2LADSDEMO2025} was designed to visualize the information captured from each monitored learner. This system enabled the following: \textit{i)} Validation of correct monitoring and synchronization: visualizing the information from the captured signals (webcam videos, screen recordings, heart rate, attention level, etc.) and ensuring their correct synchronization. \textit{ii)} Statistical analysis of the data:  comparing learner performance (pretest vs. posttest scores), analyzing signals across activities (e.g., identifying which activities had higher attention levels or heart rates), and correlating biometric signals.
The overall architecture, including the platforms and tools used to gather IMPROVE dataset, is summarized in Fig.~\ref{fig:Platforms}.

\subsubsection*{Data Processing}

\paragraph{Facial Video Processing.} 

The videos from the RealSense RGB camera were processed to extract the learner’s facial position, facial landmarks, and 3D head pose for each frame. Three state-of-the-art methods based on Convolutional Neural Networks (CNNs) were employed:
\begin{itemize}

\item Facial Detection: The facial detection module employed MediaPipe’s BlazeFace\cite{bazarevsky2019blazeface} model (full-range version) for 2D facial image detection. This module is based on CNNs specifically designed to significantly reduce computational cost and inference time without compromising accuracy. BlazeFace integrates a lightweight feature extraction architecture inspired by MobileNetV1/V2\cite{sandler2018mobilenetv2}, leveraging their efficiency and speed. Specifically, the model’s architecture applies advanced convolution techniques, similar to those employed in networks like CenterNet\cite{duan2019centernet}, with a customized encoder. The detector was trained on large private datasets containing geographically diverse facial detection samples.

\item Head Pose Detection: A real-time head pose detector based on the WHENet\cite{zhou2020whenet} architecture was employed. The Euler angles —pitch (vertical), yaw (horizontal), and roll (longitudinal)— were estimated from 2D head images, allowing the 3D orientation of the head to be deduced from 2D facial images. This architecture is based on fully connected CNNs for angle classification. Specifically, EfficientNet\cite{tan2019efficientnet} was used as the backbone for feature extraction. The architecture consists of several layers, starting with convolutional layers that capture spatial features of the image. These convolutional layers were followed by pooling layers.
A series of fully connected dense layers interprets the extracted features and performs the classification and regression of the Euler angles. In the classification process, each of the angles (yaw, pitch, and roll) was classified into bins, while regression was used to accurately predict the Euler angle values within those bins.
The head pose detector was trained on the 300W-LP\cite{zhu2016face} and CMU\cite{joo2015panoptic}.

\item Facial Landmark Detection: The MediaPipe library was employed to estimate 478 3D facial landmarks in real time from facial images. This enables the extraction of relevant information from facial regions such as the eyes, forehead, mouth, and more. The model is based on a CNN, similar to MobileNetV2\cite{sandler2018mobilenetv2}, but with custom blocks designed for real-time performance.
The model was evaluated on a private dataset comprising 1,700 examples uniformly distributed across 17 geographical subregions. It achieved a normalized mean error of 2.62\%, calculated relative to the inter-ocular distance, with similar error rates observed for each subregion. Consequently, each video is accompanied by the 478 3D facial landmarks for every frame, along with detailed information about the estimation of each landmark, including its visibility and presence in the image.
\end{itemize}
\paragraph{Eye Tracker Signal Processing.} 
The data obtained from the eye tracker was processed to extract relevant features such as fixations and saccades. On average, Tobii Pro Fusion captures one sample with gaze coordinates every 8.3 ms. These samples can be processed over the entire video duration to obtain relevant visual attention features, such as fixations and saccades. For this purpose, one of the built-in filters provided by the Tobii Pro Lab software was selected, specifically the I-VT (Fixation) filter.
We selected this filter because it is suitable for our study, where head movements are minimal and the predominant eye movements are fixations and saccades. The filter classifies gaze points as either fixations or saccades based on their velocity: points with speeds below 30 degrees per second are grouped into a single fixation. Some characteristics of this filter include: \textit{i)} Noise reduction: A moving average is applied every 3 samples. \textit{ii)} Velocity calculation: Velocity is calculated in 20 ms windows. \textit{iii)} Minimum fixation duration: Fixations shorter than 60 ms are discarded. iv) Merging adjacent fixations: Adjacent fixations are merged if the interval between them is 75 ms or less and the angular distance is 0.5 degrees or less.

 \begin{figure*}[t]
    \centering
    \includegraphics[width=\textwidth]{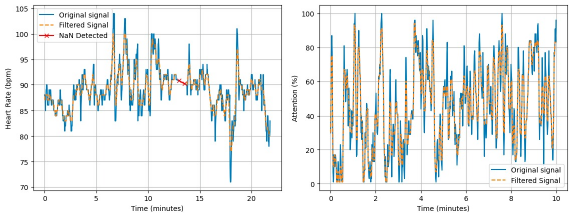} 
    \caption{Example of filtered signals in IMPROVE: The left graph displays the raw heart rate signal and the same signal filtered using a moving average filter (with a 15-second window). The right graph shows the raw attention signal derived from the EEG and filtered using a median filter (with a 5-second window). In both graphs, the original signals are represented by a solid blue line, while the filtered signals are shown with a solid yellow line.}
    \label{fig:Filter_signals}
\end{figure*}

 \paragraph{EEG Signal Processing.} The data obtained from the EEG headband were filtered to mitigate potential errors, such as signal fluctuations caused by sudden participant movements, ensuring more accurate EEG recordings. The signals were processed using a median filter with a window size of 5, effectively reducing noise from artifacts, such as participant movements, eyeblinks, and other sources, without distorting the peaks or extreme values in the signal. Additionally, interpolation was applied to data losses shorter than 5 seconds, while longer data gaps were marked as missing. Both the raw and filtered signals are available in the dataset.

\paragraph{Smartwatch Signal Processing.} The heart rate data were also filtered to remove minor fluctuations and smooth the signal. Specifically, a moving average filter with a 15-second window was applied, as the pulse signal in this context is expected to change slowly, with beat-to-beat variations typically not being abrupt at the sampling rate of smartwatches. The signal obtained from the smartwatch’s accelerometer was filtered using a fourth-order Butterworth low-pass filter with a cutoff frequency of 15 Hz. This approach was used to eliminate high-frequency noise, such as vibrations or unintentional movements, while preserving the rapid and relevant movements during the learner’s interactions with the mouse. For the gyroscope signal, a similar Butterworth low-pass filter was applied, but with a cutoff frequency of 10 Hz, which allowed for the capture of rapid wrist movements while eliminating potential high-frequency artifacts. Both the raw and filtered data are available in the dataset. Fig.~\ref{fig:Filter_signals} presents examples of filtered signals from the IMPROVE dataset. 

\section*{Data Records}

The IMPROVE dataset is available from the Science Data Bank\cite{Daza2024IMPROVE} (\url{https://doi.org/10.57760/sciencedb.14565}) and GitHub (\url{https://github.com/BiDAlab/IMPROVE}). The dataset comprises 2.83 terabytes of information, including biometric data, physiological signals, videos, supervisor-labeled events related to mobile phone usage, and more. This data was collected from the edBB and edX platforms and the LOGGE tool, and includes both raw and processed signals. The dataset is organized into three main folders, each corresponding to a different group based on mobile phone usage (see Subsection Protocol for more details). Within each of these folders, there are 40 main subfolders, each labeled with the learner’s ID, containing the monitored data for each learner during the learning session. Each of these  subfolders is further divided into four categories: 'edBB', 'edBB Processed Data', 'edX,' and 'LOGGE,' corresponding to the information collected from each platform and tool. The structure and contents of the dataset are summarised in the following tables: Table~\ref{tab:files_processed_edBB}, Table~\ref{tab:files_logge_edX}, and Table~\ref{tab:files_edbb}. The processed signals and videos obtained through the edBB platform are stored in the 'edBB Processed Data' folder and are described in Table~\ref{tab:files_processed_edBB}.   Table~\ref{tab:files_logge_edX} lists the files and descriptions from the edX platform and the LOGGE tool. Additionally, all files from the edBB platform and their descriptions are summarized in Table~\ref{tab:files_edbb}.

\begin{table}[h!]
\centering
\setlength{\tabcolsep}{6pt} 
\renewcommand{\arraystretch}{1.2} 
\begin{tabular}{|p{2.2cm}|p{3cm}|p{3cm}|p{1.2cm}|p{6cm}|}
\hline
\textbf{Folder Name} & \textbf{Files} & \textbf{Description} & \textbf{Format} & \textbf{File Contents} \\
\hline
\textbf{Video} & box\newline head\_pose\newline landmarks & Processed video data: Bounding boxes, facial landmarks, and head pose of all users in the RealSense video & .csv\newline .csv\newline .csv & Bounding box  is formatted as [x, y, width, height], 478 landmarks per detected face, Euler angles  (yaw, pitch, and roll) in degrees, frames \\

\hline
\textbf{MindWave} & filter\_ATT\newline filter\_MED\newline filter\_BANDPOWER & EEG band signals filtered with a median filter & .csv\newline .csv\newline .csv & Attention (0-100), meditation (0-100), blink strength, 5 EEG waves (dB), local date and time \\

\hline
\textbf{Smartwatch} & filter\_ACC \newline filter\_Gyro \newline filter\_Heart & Huawei Watch 2 signals filtered using a Butterworth low-pass filter or a moving average filter & .txt\newline .txt \newline .txt &  Acceleration on X, Y, Z axes (m/s\(^2\)), angular velocity on X, Y, Z axes (\(^\circ/s\)), heart rate (bpm), UNIX timestamp \\
\hline

\textbf{Fitbit} & Heart  & Heart rate signal from the FitBit Sense  filtered with a Butterworth low-pass filter & .csv & Heart rate (bpm), timestamp \\
\hline

\end{tabular}
\caption{Descriptions of the processed edBB files included in the edBB Processed Data folder.}
\label{tab:files_processed_edBB}
\end{table}

\begin{table}[h!]
\centering
\setlength{\tabcolsep}{6pt} 
\renewcommand{\arraystretch}{1.2} 
\begin{tabular}{|p{2.2cm}|p{3cm}|p{3cm}|p{1.2cm}|p{6cm}|}
\hline
\textbf{Folder Name} & \textbf{Files} & \textbf{Description} & \textbf{Format} & \textbf{File Contents} \\
\hline
\textbf{Logge} & file\_events\_LS\newline file\_marks\newline file\_position\newline user\_data \newline file\_prestest & Activities and events logs & .csv\newline .csv\newline .csv\newline .csv \newline .csv & Key name for the activity/event, start timestamp, end timestamp, marks of assignments, mobile phone position in table, medical issues, group, pretest mark, posttest mark, self-reported stress before and after the session, self-reported distraction before and after the session,  self-reported difficulty of session, self-reported performance\\
\hline
\textbf{edX} & file & edX logs & .csv & Action name, context, session, time, event type \\
\hline
\end{tabular}
\caption{Description of files collected by the edX platform and LOGGE tool, included in the edX and LOGGE folders.}
\label{tab:files_logge_edX}
\end{table}

\begin{table}[p]
\centering
\setlength{\tabcolsep}{6pt} 
\renewcommand{\arraystretch}{1} 
\begin{tabular}{|p{2.2cm}|p{3cm}|p{3cm}|p{1.2cm}|p{6cm}|}
\hline
\textbf{Folder Name} & \textbf{Files} & \textbf{Description} & \textbf{Format} & \textbf{File Contents} \\
\hline
\textbf{MouseCapture} & Mouse\_Event\newline Mouse\_Move \newline Mouse\_Dragged \newline Mouse\_Wheel \newline Mouse\_Click & Mouse click events and movement tracking & .csv\newline .csv \newline .csv \newline .csv \newline .csv& Press/Release events, mouse button pressed, position (pixels), drag-and-drop detection, scroll movement, click count, Unix timestamp, frame from the webcams \\
\hline

\textbf{Keyboard} & Keyboard\_Capture\newline Keyboard\_Key\_Typed & Keyboard data capture and typed Unicode characters & .csv\newline .csv & Press/Release events, key code, native key code, associated native Unicode character,  Unix timestamp, frame from the webcams\\
\hline

\textbf{MindWave} & file\_ATT\newline file\_MED \newline file\_BLINK \newline  file\_BANDPOWER & EEG data & .csv\newline .csv \newline .csv \newline .csv & Attention (0-100), meditation (0-100), blink strength, 5 EEG waves (dB), local date and time \\
\hline
\textbf{Smartwatch} & ACC \newline Gyro \newline Heart & Huawei Watch 2 data & .txt\newline .txt \newline .txt &  Acceleration on X, Y, Z axes (m/s\(^2\)), angular velocity on X, Y, Z axes (\(^\circ/s\)), heart rate (bpm), UNIX timestamp \\
\hline

\textbf{Fitbit} & Heart \newline EDA & FitBit Sense data & .csv\newline .csv & Skin conductance levels, heart rate (bpm), timestamp \\
\hline
\textbf{Eye Tracker} & Learner ID & All information obtained from the eye tracker & .tsv & Recording timestamp (\textmu{}s), recording date (local date of the recording), recording start time (local start time), recording resolution (pixels), gaze point (pixels), pupil diameter (mm), eye movement type, validity \\
\hline
\textbf{VideoMonitor} & monitor\newline Videomonitor & Screen recording  & .csv\newline .mp4 &   Frame and timestamps recorded in both UNIX and local time \\
\hline
\textbf{VideoWebcam} & webcam\newline WebcamCapture & Side webcam recording & .csv\newline .mp4 & Frame number and timestamps recorded in both UNIX and local time \\
\hline
\textbf{VideoWebcam2} & webcam\newline WebcamCapture & Overhead webcam recording & .csv\newline .mp4 & Frame number and timestamps recorded in both UNIX and local time \\
\hline
\textbf{RealSense} & Time\newline Color \newline Depth \newline Left\_Infrared \newline Right\_Infrared & Front-facing RGB, NIR, and depth cameras & .csv\newline .mp4 \newline .mp4 \newline .mp4 \newline .mp4 & Frame number and timestamps recorded in both UNIX and local time for each camera \\
\hline

\textbf{SoundCapture} & Record & Session audio recording & .wav & -- \\
\hline

\textbf{PCInformation} & PCInformationCapture & PC information & .csv & Computer Name, IP address, MAC address, operating system and version, system architecture, main HDD, free HDD, screen resolution, keyboard language \\
\hline
\textbf{StudentData} & StudentData & Session and learner information & .csv & Learner ID, session number, session start/end time (local time), session completion status indicator \\
\hline
\end{tabular}
\caption{Description of files collected by the edBB platform and contained in the edBB folder.}
\label{tab:files_edbb}
\end{table}

\clearpage

\section*{Technical Validation}

For the technical validation of the dataset, we conducted five analyses, organised into three subsections.
The first subsection, Mobile Phone Event Labelling Accuracy, evaluates the precision of mobile phone event annotations and the changes in physiological and biometric signals during these events. It includes:
(i) Semi-supervised Event Data Curation, and
(ii) Statistical Validation of Signal Changes During Mobile Phone Usage Events.
The second subsection, Signal Quality Assessment, assesses the overall quality and consistency of the signals recorded by the sensors through:
(iii) Statistical Distributions of Physiological and Biometric Signals, and
(iv) Verification of Signals and Video Recordings with M2LADS.
Finally, the third subsection, Inter-Device Comparisons, analyses the consistency of heart rate estimations across different smartwatch devices:
(v) Heart Rate Data Interoperability Between Devices.

\subsection*{Mobile Phone Event Labelling Accuracy}
\vspace{2mm}
\subsubsection*{i) Semi-supervised Event Data Curation.}  
\begin{figure*}[t]
    \centering
    \includegraphics[width=\textwidth]{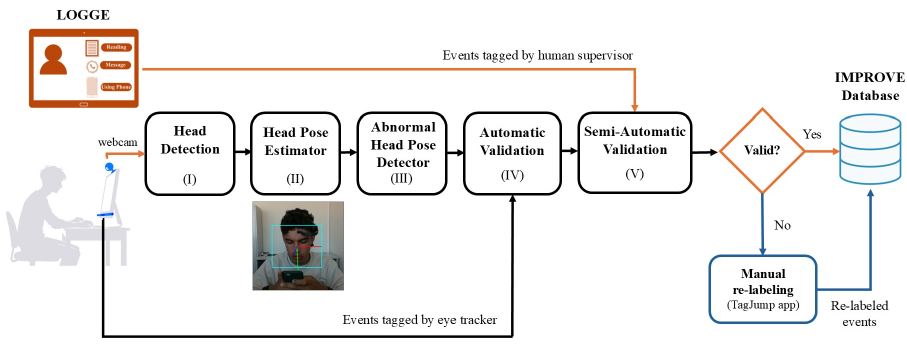} 
    \caption{Diagram of the proposed semi-supervised method for curating mobile phone usage event data. The participant provided consent for the publication of this image. 
 }
    \label{fig:Diagram}
\end{figure*}

A semi-supervised method was employed for the data curation of the mobile phone usage events. To this end, we used the eye tracker data and head pose estimation through video (see Fig. \ref{fig:Diagram}) for the automatic detection of mobile phone usage events. These automatically annotated events were validated against the events labeled by the human supervisor, who manually labeled the start and end of each event.

\begin{figure*}[t]
    \centering
    \includegraphics[width=\textwidth]{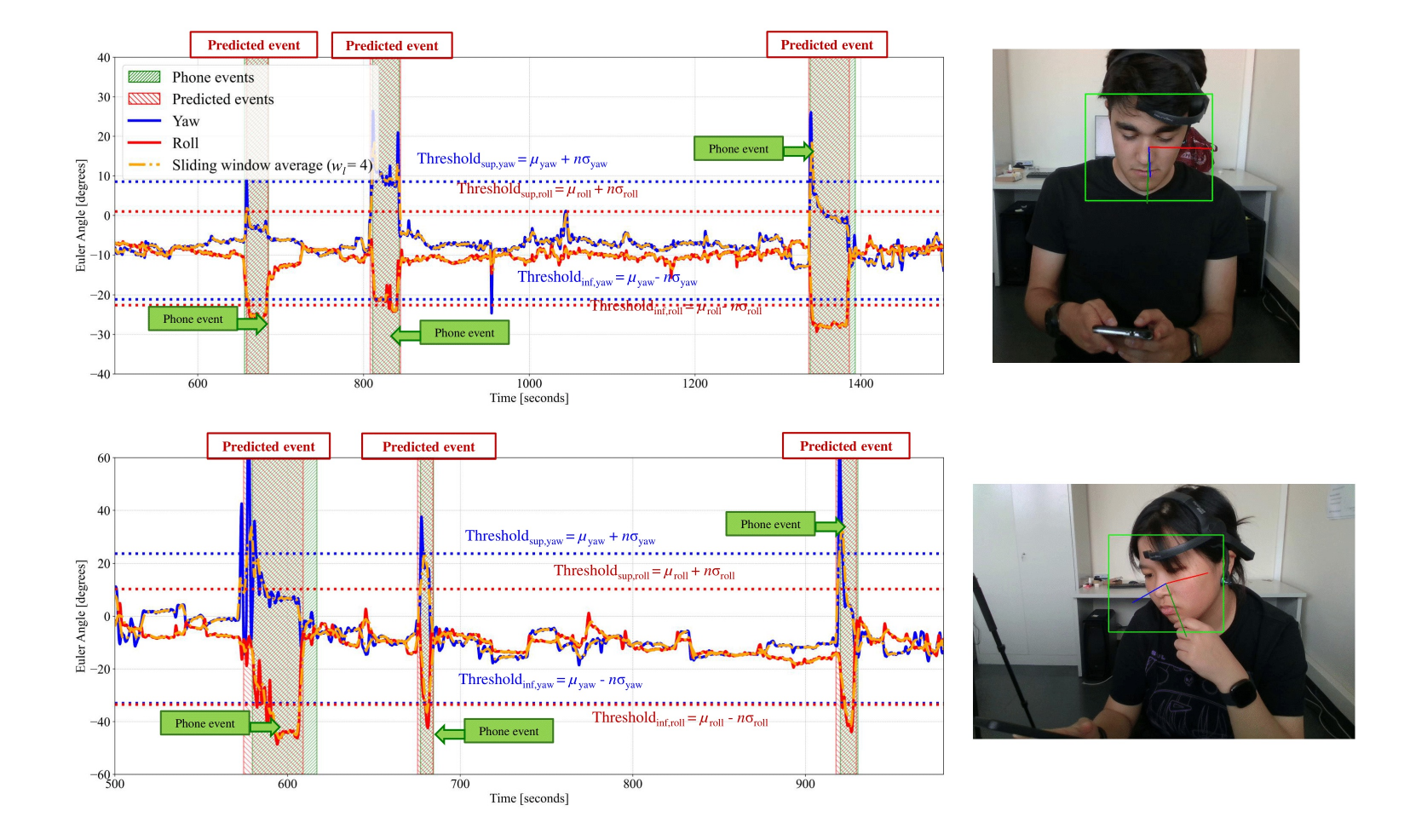} 
    \caption{Example of the abnormal head pose detector for two students. The graphs show the Euler angles obtained during the session, along with the average for each angle in a window \( W_l = 4s \). The set thresholds for each angle are also shown; values exceeding these thresholds are marked as events. The detector exhibits a high overlap with the manually labeled events. The images show the students’ pose while using a mobile phone. Yaw is represented by the solid blue line, and roll by the solid red line. The temporal window average is indicated by the dashed yellow line. Predicted event labels are framed in red, while manually labeled events are framed in green. Thresholds are represented by dotted lines with the corresponding color for each angle. The participants provided consent for the publication of these images.
  }
    \label{fig:Pose}
\end{figure*}

The semi-supervised data curation method processes video data from the RGB webcam to detect head pose changes and moments when the learner was not looking at the computer, which with high probability, could correspond to mobile phone usage (see Fig. \ref{fig:Diagram}). To process the videos and detect head pose changes, we used two state-of-the-art modules based on CNNs, which are described in more detail in the Data Processing section. 
\begin{itemize}
\item In the first module (I), the video frames were processed to detect the learner’s face by selecting the most centered and widest bounding box for each frame, using MediaPipe’s BlazeFace facial detector \cite{bazarevsky2019blazeface}. 
\item In the second module (II), the WHENet head pose estimator \cite{zhou2020whenet} was applied to obtain the Euler angles (pitch, roll, and yaw).  

\item In the third module (III), the method presented by Becerra et al. \cite{Becerra2024} was followed to detect abnormal head pose events. The process involves calculating the Euler angles (pitch, yaw, and roll) for each video frame. A sliding window ($W_l$) methodology was then applied to compute the average angles within each temporal window. An event was detected when the local average of an angle in a window exceeded a predefined threshold, which was determined by a significant deviation from the global average. This method ensures that only meaningful head pose changes are flagged as events, minimizing false positives from minor movements.
The method consists of the following steps: \begin{enumerate}  \item The mean ($\mu$) and standard deviation ($\sigma$) for each Euler angle were calculated for all sessions. \item For each temporal window $W_l$, the average $\mu$ of the Euler angles was calculated, providing a representative value for pitch, yaw, and roll within that specific time frame.
\item A temporal window $W_l$ was labeled as an event when the $(\mu_{W_l})$ exceeded a predefined threshold, which was set based on a significant deviation from the global mean ($\mu$) of the Euler angles calculated over all sessions. The following equation shows the threshold: $\left|\mu_{W_l}^i - \mu^i\right| > n\sigma^i$, \noindent where $i$ represents the angle (yaw, pitch, or roll) and  $n$  is the threshold factor that defines the sensitivity of the event detection, determining how many standard deviations away from the mean a particular value must be to be considered an event.


\item All overlapping events were merged.

\end{enumerate}

Figure \ref{fig:Pose} shows the detection of abnormal head pose events applied in module (III), for two learners from the IMPROVE dataset.

\item In the fourth module (IV), information from the eye tracker was used to discard abnormal head pose false detections in cases where the learner continued to look at the computer screen.

\item In the fifth module (V), the information from the LOGGE tool was used to determine when mobile phone usage events were labeled. This information helped select only the predicted events from our abnormal head pose detector that occurred within the time window  $W_s$ indicated by the supervisors, including a buffer \( \Delta t = 10 \, \text{s} \) before and after the event. The assumption is that manual labeling was generally accurate but involved slight delays when marking the start and end of each event.
The conditions for maintaining the supervisor’s labeling were as follows: the abnormal head pose detector must report the same number of events within  \( W_s \pm \Delta t \) as those recorded by LOGGE, and the overlap between the predicted and manually labeled events must have a delay of less than 4 seconds. If these conditions are not met, manual re-labeling is required using the TagJump application we developed. TagJump allows displaying the video of the session at the start and end times of events reported by LOGGE and those predicted by the abnormal head pose detector. It also enables quick navigation through nearby frames to facilitate re-labeling. The protocol of this application first shows the LOGGE-labeled event, and if the event is not found within the first 4 seconds, the event predicted by the detector is displayed. This approach simplifies the supervisor’s labeling process.
\end{itemize}

\begin{figure*}[t]
    \centering
    \includegraphics[width=\textwidth]{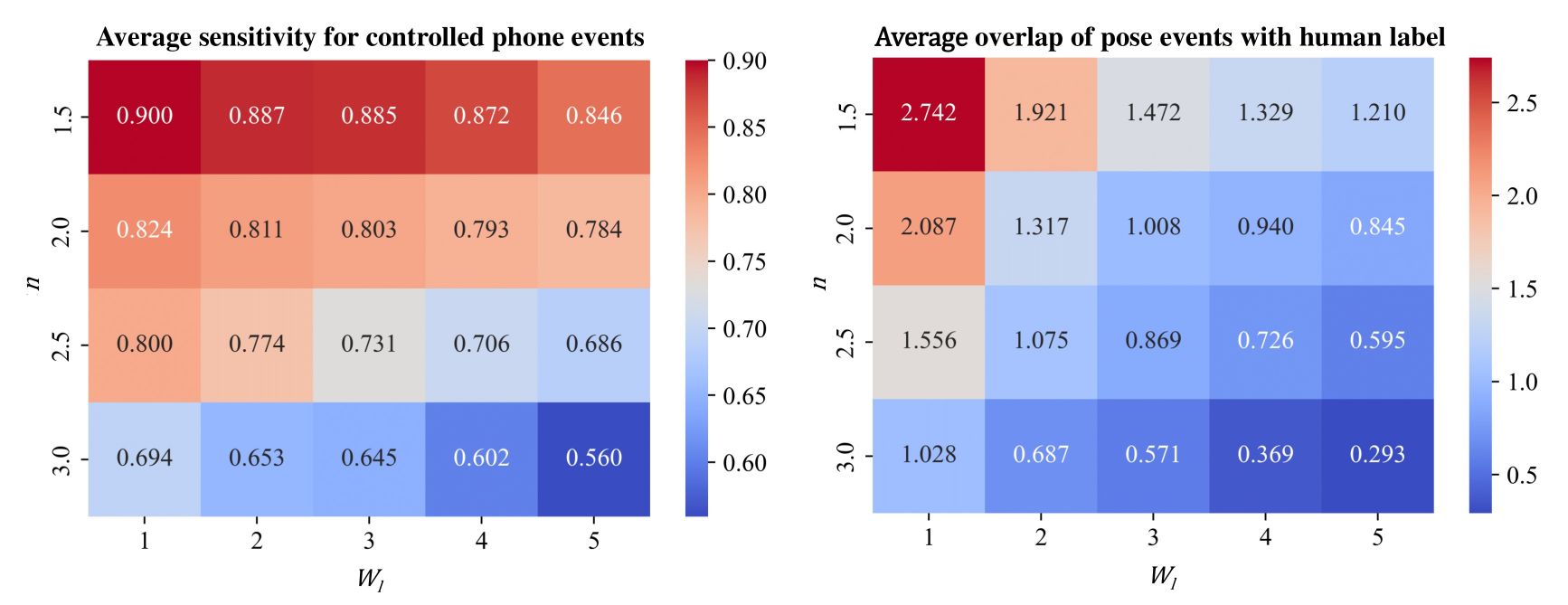} 
    \caption{Average event detection sensitivity obtained by the abnormal head pose detection approach to detect mobile phone usage across all users (left). Average number of detected events by the same approach across all users within the window \( W_s \pm \Delta t \) (right). The x and y axes in both heatmaps represent the tested parameters \( W_l \) and \( n \), respectively, which were varied to analyze their effect on detection performance.
 }
    \label{fig:Heatmaps}
\end{figure*}

To fine-tune the parameters $W_l$ and $n$, a preliminary study was conducted using data labeled by the LOGGE tool (see the previous subsection Data, Platforms, and Sensors), specifically focusing on the controlled events, which were presumed to have been labeled with greater precision by the supervisors. The goal was to optimize the parameters to maximize the detection of mobile phone usage events while minimizing false positives, ensuring that the predicted windows closely matched those labeled by the supervisors. Fig. \ref{fig:Heatmaps} presents two heatmaps for controlled events. The first heatmap illustrates the event detection accuracy, while the second shows the average number of detected events within the windows  \( W_s \pm \Delta t \). Both are dependent on the selected parameters and the temporal window \( W_s \pm \Delta t \). Based on the data analysis, we selected $W_l=4s$ and $n=2$ as the optimal parameters. However, this choice is not straightforward, and other combinations of parameters could also be valid depending on the specific use case. 

Following this protocol, 80\% of the LOGGE events were re-labeled. Only 20\% had a delay of less than 4 seconds. 
The LOGGE labels and those detected by the abnormal pose detector complemented each other. In 24.80\% of the events, when one of the labels had a delay of more than 4 seconds, the other label had a delay of less than 4 seconds. This complementarity between the two labeling systems facilitated faster re-labeling with the TagJump application.
27.2\% of the re-labeled events showed a difference of less than 3 seconds, 15.2\% had a difference between 3 and 4 seconds, while 57.6\% showed a delay of more than 4 seconds compared to the LOGGE reports.

The re-labeled events were checked to determine whether they improved the event detection accuracy of the abnormal head pose detector. For controlled events, accuracy improved by approximately 3\%, and for uncontrolled events, by around 5\%, demonstrating that the detector becomes even more accurate with proper re-labeling.

\subsubsection*{ii) Statistical Validation of Signal Changes Due to Mobile Phone Usage:} The absolute mean and standard deviation of the biometric and physiological signals were calculated both during mobile phone usage events and when the mobile phone was not in use. A statistical analysis was then conducted to determine whether significant changes occurred in the signals during mobile phone usage. The results in Table \ref{tb:phono_no_phono} indicate that, according to the Student’s t-test, almost all signals have p-values below 0.05, suggesting a low likelihood that the observed differences between phone and no-phone conditions are due to chance. Overall, these findings suggest that significant impacts on these variables are caused by mobile phone usage.

Notably, the accelerometer and gyroscope signals exhibit highly significant differences between the phone and no-phone conditions, as expected due to the increased movement associated with mobile phone use, which contrasts with typical movements during an online session. Heart rate also shows a significant difference, indicating that it is affected by mobile phone use. Most EEG signals, including attention, also show significant differences. However, the meditation, beta, and theta signals do not exhibit statistically significant differences in their mean values. Theta waves are related to relaxation and wakeful states, so it is unsurprising that the Meditation signal shows no significant differences, suggesting that relaxation or meditation is not notably affected by mobile phone usage. However, further research is required to confirm this.

Beta waves are associated with focus, normal waking consciousness, and alertness \cite{lin2018mental, chen2017assessing}. For instance, beta waves occur during problem-solving, decision-making, and anxious thinking. Nevertheless, beta did not show statistically significant differences in this study.

\begin{figure}
    \centering
    \includegraphics[width=\linewidth]{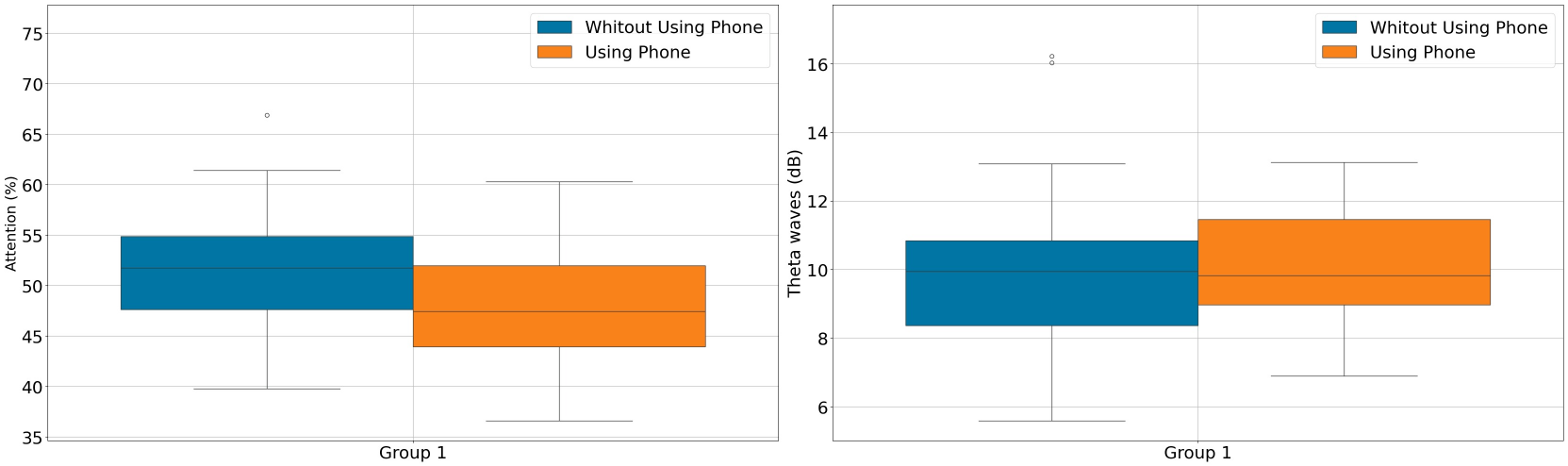}
    \caption{Boxplots comparing mobile phone usage and non-usage conditions for attention levels (left) and theta wave activity (right).}
    \label{fig:Box_plot}
\end{figure}

Figure \ref{fig:Box_plot} presents boxplots for attention and theta waves across two experimental conditions: "Without Using Phone" and "Using Phone." Attention levels tend to be lower when the mobile phone is used, while the range of values is broader when the phone is not in use. In contrast, theta waves show a slight increase during mobile phone usage.

In summary, most signals are affected by mobile phone usage, validating the correct labeling of these events and supporting the potential to study the effects of mobile phone use through the IMPROVE dataset.

\begin{table}[htbp]
    \centering
    \begin{tabular}{|l|l|c|c|c|c|c|}
    \cline{3-7}
    \multicolumn{2}{c|}{} & \multicolumn{3}{c|}{\textbf{Mean (Std)}} & \multicolumn{2}{c|}{\textbf{Statistical Tests}} \\
    \hline
    \textbf{Sensors} & \textbf{Variable (Unit)} & \textbf{No Phone} & \textbf{Phone} & \textbf{All Sessions} & \textbf{t} & \textbf{p-value} \\
    \hline
    \multirow{6}{*}{\centering Smartwatch} & Heart Rate (bpm) & 82.68 (10.3) & 84.03 (10.29) & 82.85 (10.25) & -2.56 & 0.0153 \\
    & Acc\(_x\) (m/s\(^2\)) & 0.74 (0.48) & 2.52 (0.99) & 0.98 (0.48) & -10.82 & 0.0000 \\
    & Acc\(_y\) (m/s\(^2\)) & 4.58 (1.55) & 8.05 (1.03) & 5.05 (1.39) & -12.15 & 0.0000 \\
    & Acc\(_z\) (m/s\(^2\)) & 8.17 (1.00) & 3.31 (1.25) & 7.51 (0.89) & 16.80 & 0.0000 \\
    & Gyro\(_x\) (°/s) & 0.05 (0.02) & 0.22 (0.09) & 0.07 (0.03) & -11.25 & 0.0000 \\
    & Gyro\(_y\) (°/s) & 0.03 (0.01) & 0.09 (0.04) & 0.04 (0.01) & -10.50 & 0.0000 \\
    & Gyro\(_z\) (°/s) & 0.04 (0.01) & 0.09 (0.04) & 0.04 (0.02) & -9.70 & 0.0000 \\
    \hline
    \multirow{7}{*}{\centering EEG Band} & Attention (0-100) & 51.28 (5.5) & 48.47 (8.05) & 50.91 (5.68) & 3.03 & 0.0048 \\
    & Meditation (0-100) & 56.33 (4.4) & 56.80 (6.51) & 56.39 (4.46) & -0.59 & 0.5626 \\
    & Alpha (dB) & 6.31 (1.72) & 6.85 (1.78) & 6.38 (1.69) & -3.57 & 0.0011 \\
    & Beta (dB) & 3.30 (0.95) & 3.45 (0.98) & 3.32 (0.94) & -2.00 & 0.0541 \\
    & Delta (dB) & 1.69 (0.43) & 1.80 (0.49) & 1.70 (0.43) & -2.35 & 0.0250 \\
    & Gamma (dB) & 1.86 (0.80) & 2.03 (0.72) & 1.89 (0.78) & -2.36 & 0.0243 \\
    & Theta (dB) & 9.99 (2.36) & 10.18 (2.13) & 10.02 (2.31) & -0.88 & 0.3839 \\
    \hline
    \end{tabular}
    \caption{Absolute mean and standard deviation for physiological, biometric, and movement signals across phone and no phone conditions, with t-values and p-values highlighting statistical significance, for group 1 in the IMPROVE dataset.}
    \label{tb:phono_no_phono}
\end{table}

\begin{table}[H]
    \centering
    \begin{tabular}{|l|l|c|c|c|c|}
    \cline{3-6}
        \multicolumn{2}{l|}{} & \multicolumn{4}{c|}{\textbf{Mean (Std)}} \\
        \hline
        \textbf{Sensors} & \textbf{Variable (Unit)} & \textbf{Group 1} & \textbf{Group 2} & \textbf{Group 3} & \textbf{Overall} \\
        \hline
        \multirow{6}{*}[0.5ex]{\centering Smartwatch} & Heart Rate (bpm) & 82.85 (10.25) & 79.82 (13.70) & 78.26 (14.30) & 80.31 (12.75) \\
        & Acc\(_x\) (m/s\(^2\)) & 0.98 (0.48) & 0.81 (0.58) & 0.75 (0.50) & 0.85 (0.52) \\
        & Acc\(_y\) (m/s\(^2\)) & 5.05 (1.39) & 4.19 (1.54) & 4.82 (1.39) & 4.69 (1.44) \\
        & Acc\(_z\) (m/s\(^2\)) & 7.51 (0.89) & 8.46 (0.94) & 8.13 (0.95) & 8.03 (0.92) \\
        & Gyro\(_x\) (°/s) & 0.07 (0.03) & 0.04 (0.03) & 0.04 (0.02) & 0.05 (0.03) \\
        & Gyro\(_y\) (°/s) & 0.04 (0.01) & 0.02 (0.01) & 0.02 (0.01) & 0.03 (0.01) \\
        & Gyro\(_z\) (°/s) & 0.04 (0.02) & 0.03 (0.02) & 0.03 (0.01) & 0.03 (0.02) \\
        \hline
        \multirow{7}{*}[0.5ex]{\centering EEG Band} & Attention (0-100) & 50.91 (5.68) & 54.98 (4.40) & 51.21 (6.54) & 52.37 (5.54) \\
        & Meditation (0-100) & 56.39 (4.46) & 54.25 (5.48) & 54.58 (5.79) & 55.07 (5.24) \\
        & Alpha (dB) & 6.38 (0.62) & 5.47 (0.61) & 6.23 (0.61) & 6.03 (0.61) \\
        & Beta (dB) & 3.32 (0.62) & 3.28 (0.58) & 3.67 (0.58) & 3.42 (0.59) \\
        & Delta (dB) & 1.70 (0.97) & 1.59 (1.00) & 1.59 (0.95) & 1.63 (0.97) \\
        & Gamma (dB) & 1.89 (0.90) & 1.97 (0.82) & 2.06 (0.75) & 1.97 (0.82) \\
        & Theta (dB) & 10.02 (0.69) & 9.21 (0.71) & 9.84 (0.71) & 9.69 (0.70) \\
        \hline
    \end{tabular}
    \caption{Absolute mean and standard deviation for physiological and biometric signals across the different groups in the IMPROVE dataset.}
    \label{tb:signals_mean_std}
\end{table}

\subsection*{Signal Quality Assessment}
\vspace{2mm}
\subsubsection*{iii) Statistical Distributions of Physiological and Biometric Signals:}

The mean and standard deviation of each biometric and physiological signal were calculated to confirm the absence of general errors and to verify that the values fell within expected ranges for a similar environment (see Table~\ref{tb:signals_mean_std}). These results were also compared with the mEBAL2 dataset \cite{daza2024mebal2}, which was obtained in a similar e-learning environment and utilized the same EEG band. The values obtained for the physiological and biometric signals were found to fall within the expected ranges for learners in an e-learning context. The average heart rate across the three groups ranged from 78.26 to 82.85 bpm, consistent with values reported in previous studies on students during virtual learning sessions \cite{hernandez2020heart}, \cite{hernandez2019edbb}, and with the normal human resting heart rate, typically between 60-100 bpm \cite{palatini1997heart}.

The inertial sensors (accelerometer and gyroscope) in the Huawei Watch 2 smartwatch, worn on the hand used to operate the mouse, provided the following observations.
The average acceleration in the X and Y axes was low, which is expected during a learning session where most of the time is characterized by either no movement or minor mouse movements. In contrast, the higher values on the Z axis were primarily attributed to the influence of gravitational force. Similar results were observed in the gyroscope, with mean values close to zero across all axes, reflecting the same reasons.

For attention and meditation, the means were 52.37 and 55.07, respectively, aligning with previous studies in e-learning environments, such as the mEBAL2 dataset \cite{daza2024mebal2, daza2024deepface}, which also report values around 50\% for attention and 53.88\% for meditation. These findings are in line with moderate-length sessions, where attention generally starts high but gradually decreases over time, stabilizing around the mean without extended periods of low attention \cite{mendoza2018effect}.
EEG signals (alpha, beta, delta, gamma, theta): The average values of the EEG frequency bands in IMPROVE are consistent and comparable to those obtained in the mEBAL2 dataset, where alpha (8.09 dB), beta (5.22 dB), delta (2.00 dB), gamma (2.59 dB), and theta (9.93 dB) are reported. In both datasets, the order of the signals according to their relative power remains constant, and the value ranges are very similar. The slight differences observed can be attributed to the varying learning activities performed in each dataset, as well as the intrinsic differences between users.

\begin{table}[h]
\centering
\renewcommand{\arraystretch}{1.2}
\begin{tabular}{
|>{\raggedright\arraybackslash}p{3cm}
|>{\raggedright\arraybackslash}p{3.5cm}
|>{\centering\arraybackslash}p{2.0cm}
|>{\centering\arraybackslash}p{2.0cm}
|>{\centering\arraybackslash}p{2.0cm}
|>{\centering\arraybackslash}p{1cm}|
}
\hline
\textbf{Sensor / Platform} & \textbf{Data Issues} 
& \textbf{\parbox{2cm}{\centering \vspace{1mm}Group 1\\(IDs)\vspace{1mm}}} 
& \textbf{\parbox{2cm}{\centering \vspace{1mm}Group 2\\(IDs)\vspace{1mm}}}  
& \textbf{\parbox{2cm}{\centering \vspace{1mm}Group 3\\(IDs)\vspace{1mm}}} 
& \textbf{Count} \\
\hline
\multicolumn{1}{|p{3cm}|}{Camera} & Corrupt depth video & 202303281 & - & - & 1 \\
\cline{2-6}
\multicolumn{1}{|p{3cm}|}{} & Corrupt left NIR video & - & 202305233 & - & 1 \\
\hline
\multicolumn{1}{|p{3cm}|}{edX Platform} & Missing edX logs & - & 202303241 202303242 & 202305185 & 3 \\
\hline
\multicolumn{1}{|p{3cm}|}{Fitbit Sense} & Missing all heart rate  & - & 202303272 202303274 & 202303273 & 3 \\
\cline{2-6}
\multicolumn{1}{|p{3cm}|}{} & Incorrect timestamp & - & 202305102 202304283 & - & 2 \\
\hline
\multicolumn{1}{|p{3cm}|}{Huawei Watch 2} & All data missing & - & - & 202304143 & 1 \\
\cline{2-6}
\multicolumn{1}{|p{3cm}|}{} & 1-minute signal loss & 202303283 202305192 & - & 202303231 & 3 \\
\hline
\multicolumn{1}{|p{3cm}|}{EEG Band} & 1-minute signal loss 
& 202303281 202304272 202306161 202304184 202305311 202306075 
& 202304202 202305103 202306141 202304115 202304181 202305092 202304183 202304282 202304284 202306055 202307271 
& 202303291 202304212 202304125 202304131 202305055 202306074 202304143 202304283 202305042 202305051 202305231 
& 28 \\
\hline
\end{tabular}
\caption{Overview of missing or corrupted data detected through the M2LADS validation, including signal losses, timestamp inconsistencies, and corrupted recordings. The information is grouped by sensor or platform and categorized according to the learners’ assigned group based on mobile phone usage.}
\label{tab:missing_data}
\end{table}

\subsubsection*{iv) Verification of Signals and Video Recordings with M2LADS } M2LADS \cite{becerra2023m2lads} was used to validate the signals and video recordings captured during the session. Through the system’s visualization and the synchronization it provides, all data were checked, and no corrupt files were reported by the system.  Table \ref{tab:missing_data} presents the results of the validation process carried out using M2LADS.

\begin{figure*}[t]
    \centering
    \includegraphics[width=\textwidth]{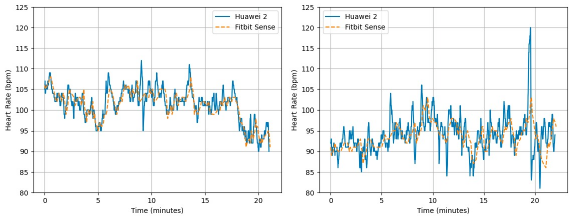} 
    \caption{Comparison of heart rate measurements between the Huawei 2 and Fitbit Sense for two students. The solid blue line represents the Huawei 2, while the orange dashed line corresponds to the Fitbit Sense.}
    \label{fig:heart rate}
\end{figure*}

\subsection*{Inter-Device Comparisons}
\vspace{2mm}
\subsubsection*{v) Heart Rate Data Interoperability between Devices:} The heart rate data from both smartwatches (Huawei Watch 2 and Fitbit Sense) were compared to ensure consistent measurements, as similar values would indicate reliable data recording. The heart rate sampling frequency differs between the two devices, with data being captured by the Huawei Watch 2 at 1 Hz, while the Fitbit Sense operates at a lower frequency of 0.2 Hz. A difference of 3.70 bpm was found between the two signals (see Fig. \ref{fig:heart rate}), which falls within the reasonable range for this type of device, typically around $\pm 3\text{-}5$ bpm \cite{hwang2019assessing}. Additionally, it should be noted that the smartwatches were worn on different wrists, with the dominant hand experiencing more movement, which may have affected heart rate estimation. This, along with the variation in sampling frequencies, may have contributed to the observed differences.

\section*{Usage Notes}

\textbf{Suggestions:} It is strongly recommended to use the pre-filtered EEG band files or apply appropriate filtering techniques using programs like Matlab or Python. In Python, the \textbf{SciPy}\cite{virtanen2020scipy} library (\url{https://scipy.org}) provides algorithms for signal processing, including various filtering algorithms. The EEG band is often affected by artifacts caused by movement and eyeblinks, so it is recommended to apply a moving average or median filter to improve the quality of the signals \cite{daza2024deepface}.

The \textbf{Pandas} library (\url{https://pandas.pydata.org/}) proved very useful for manipulating and analyzing large datasets, while \textbf{NumPy}\cite{harris2020array} (\url{https://numpy.org}) was used for processing various types of signals and videos. For training deep learning models or neural networks, \textbf{TensorFlow}\cite{abadi2016tensorflow} (\url{https://www.tensorflow.org/}) and \textbf{PyTorch}\cite{paszke2019pytorch} (\url{https://pytorch.org/}) are two of the most powerful and flexible platforms for this purpose, and were essential for running the models used in video processing.

For video processing, the \textbf{Dlib}\cite{king2009dlib} library (\url{www.dlib.net}), which includes pre-trained models for facial detection, facial landmark estimation, and more, was used. This library integrates well with \textbf{OpenCV} (\url{https://opencv.org/}), which facilitates video reading, filtering, distortion correction, and also provides additional pre-trained models. This combination was used to process the videos, alongside \textbf{MediaPipe}\cite{lugaresi2019mediapipe} (\url{https://github.com/google-ai-edge/mediapipe?tab=readme-ov-file}), a library that specializes in fast models for face detection, facial landmarks, image segmentation, and more. MediaPipe was employed to extract the facial bounding boxes and landmarks available in IMPROVE dataset. The use of \textbf{RetinaFace}\cite{deng2020retinaface}  (\url{https://github.com/serengil/retinaface}) is also recommended for facial detection with landmark localization. 

 \textbf{WHENet}\cite{zhou2020whenet} (\url{www.github.com/Ascend-Research/HeadPoseEstimation-WHENet}) is a fast and efficient method for head pose estimation, which was used to obtain the Euler angle labels in the IMPROVE dataset. Another effective alternative for head pose estimation is \textbf{RealHePoNet} \cite{berral2021realheponet}  (\url{https://github.com/rafabs97/headpose_final}).

For processing eye-tracker data, the official software \textbf{Tobii Pro Lab} (\url{https://www.tobii.com/products/software/behavior-research-software/tobii-pro-lab}) is recommended, which was used in IMPROVE. Another interesting solution is \textbf{PsychoPy}\cite{peirce2007psychopy} (\url{https://www.psychopy.org/}), a free tool offering support for the creation and execution of experiments involving eye tracking.

\textbf{Data Access}: The IMPROVE dataset is available for research use only, including both academic and legitimate commercial research and development, provided that the data are not redistributed in any form (e.g., original files, encrypted files, or extracted features). Due to the inclusion of biometric signals and indirect identifiers (e.g., gender, age), access to the dataset is restricted and subject to the signing of a Data Usage Agreement (DUA). Data access will be granted to any applicant who agrees to the terms and conditions outlined in the DUA. Researchers can request access through two available methods. First, the DUA is available on the project’s GitHub (\url{https://github.com/BiDAlab/IMPROVE}); a signed and scanned copy must be emailed to atvs@uam.es, following the instructions provided in the repository. Alternatively, access requests can be submitted via the Science Data Bank\cite{Daza2024IMPROVE} (\url{https://doi.org/10.57760/sciencedb.14565}), where detailed application procedures are provided.

\section*{Code Availability}
Example code is available on GitHub (\url{https://github.com/BiDAlab/IMPROVE}) for filtering physiological and biometric signals, as well as for detecting pose events.  The repository includes:

\begin{itemize}
    \item A Python script for detecting abnormal head poses using previously extracted Euler angles. These angles are obtained through facial detection with MediaPipe   and head pose estimation with WHENet. For full functionality, users should install the corresponding dependencies from the official GitHub repositories of these tools.

    \item MATLAB scripts for filtering EEG, inertial, and heart rate signals. These scripts replicate the processing applied during dataset preparation, including median filtering, moving average smoothing, and Butterworth low-pass filtering.
\end{itemize}

The provided code is intended to support the research community in processing the IMPROVE dataset. These example scripts were used to process the raw data included in the dataset release and can serve as a reference for reproducing or extending the preprocessing pipeline.

\section*{Acknowledgements}

This work has been supported by projects: HumanCAIC (TED2021-131787B-I00 MICINN), BIO-PROCTORING (GNOSS Program, Agreement Ministerio de Defensa-UAM-FUAM dated 29-03-2022), SNOLA (RED2022-134284-T), e-Madrid-CM (S2018/TCS4307), and M2RAI (PID2024-160053OB-I00 MICIU/FEDER). Research partially funded by the Autonomous Community of Madrid. A. Morales is supported by the Madrid Government (Comunidad de Madrid-Spain) under the Multiannual Agreement with Universidad Autónoma de Madrid in the line of Excellence for the University Teaching Staff in the context of the V PRICIT (Regional Programme of Research and Technological Innovation).

\section*{Author Contributions Statement}

\textbf{R.D.:} Conceptualization, Data Collection, Data Curation, Data Analysis, Investigation, Software, Technical Validation, Writing—Original Draft, Writing—Review \& Editing.
\textbf{A.B.:}  Data Collection, Data Curation, Software, Data Analysis, Investigation, Writing—Review \& Editing.
\textbf{R.C.:} Funding Acquisition, Investigation, Supervision, Writing—Review \& Editing.
\textbf{J.F.:} Funding Acquisition, Supervision, Writing—Review \& Editing.
\textbf{A.M.:} Project Administration, Funding Acquisition, Supervision, Writing—Review \& Editing.

\section*{Competing interests}
The authors declare no competing interests.

\bibliography{sample}

\begin{thebibliography}{10}
\urlstyle{rm}
\expandafter\ifx\csname url\endcsname\relax
  \def\url#1{\texttt{#1}}\fi
\expandafter\ifx\csname urlprefix\endcsname\relax\def\urlprefix{URL }\fi
\expandafter\ifx\csname doiprefix\endcsname\relax\def\doiprefix{DOI: }\fi
\providecommand{\bibinfo}[2]{#2}
\providecommand{\eprint}[2][]{\url{#2}}

\bibitem{daza2023edbb}
\bibinfo{author}{Daza, R.} \emph{et~al.}
\newblock \bibinfo{title}{{{edBB-Demo: Biometrics and Behavior Analysis for Online Educational Platforms}}}.
\newblock In \emph{\bibinfo{booktitle}{Proc. AAAI Conf. on Artificial Intelligence (Demonstration)}}, \url{10.1609/aaai.v37i13.27066} (\bibinfo{year}{2023}).

\bibitem{GMI_eLearning2024}
\bibinfo{author}{{Global Market Insights}}.
\newblock \bibinfo{title}{{E-learning Market Size}}.
\newblock \bibinfo{howpublished}{\url{https://www.gminsights.com/industry-analysis/elearning-market-size}} (\bibinfo{year}{2023}).

\bibitem{ma2019investigating}
\bibinfo{author}{Ma, L.} \& \bibinfo{author}{Lee, C.~S.}
\newblock \bibinfo{journal}{\bibinfo{title}{{Investigating the Adoption of MOOCs: A Technology--User--Environment Perspective}}}.
\newblock {\emph{\JournalTitle{Journal of Computer Assisted Learning}}} \textbf{\bibinfo{volume}{35}}, \bibinfo{pages}{89--98}, \url{10.1111/jcal.12314} (\bibinfo{year}{2019}).

\bibitem{tindell2012use}
\bibinfo{author}{Tindell, D.~R.} \& \bibinfo{author}{Bohlander, R.~W.}
\newblock \bibinfo{journal}{\bibinfo{title}{{The Use and Abuse of Cell Phones and Text Messaging in the Classroom: A Survey of College Students}}}.
\newblock {\emph{\JournalTitle{College teaching}}} \textbf{\bibinfo{volume}{60}}, \bibinfo{pages}{1--9}, \url{10.1080/87567555.2011.604802} (\bibinfo{year}{2012}).

\bibitem{huey2023impact}
\bibinfo{author}{Huey, M.} \& \bibinfo{author}{Giguere, D.}
\newblock \bibinfo{journal}{\bibinfo{title}{{The Impact of Smartphone Use on Course Comprehension and Psychological Well-being in the College Classroom}}}.
\newblock {\emph{\JournalTitle{Innovative Higher Education}}} \textbf{\bibinfo{volume}{48}}, \bibinfo{pages}{527--537}, \url{10.1007/s10755-022-09638-1} (\bibinfo{year}{2023}).

\bibitem{andrews2015beyond}
\bibinfo{author}{Andrews, S.}, \bibinfo{author}{Ellis, D.~A.}, \bibinfo{author}{Shaw, H.} \& \bibinfo{author}{Piwek, L.}
\newblock \bibinfo{journal}{\bibinfo{title}{{Beyond Self-report: Tools to Compare Estimated and Real-world Smartphone Use}}}.
\newblock {\emph{\JournalTitle{PloS One}}} \textbf{\bibinfo{volume}{10}}, \bibinfo{pages}{e0139004}, \url{10.1371/journal.pone.0139004} (\bibinfo{year}{2015}).

\bibitem{smith2011americans}
\bibinfo{author}{Smith, A.}
\newblock \bibinfo{title}{{Americans and Text Messaging}}.
\newblock \bibinfo{howpublished}{\url{https://www.pewresearch.org/internet/2011/09/19/americans-and-text-messaging/}} (\bibinfo{year}{2011}).

\bibitem{daza2024deepface}
\bibinfo{author}{Daza, R.} \emph{et~al.}
\newblock \bibinfo{journal}{\bibinfo{title}{{DeepFace-Attention: Multimodal Face Biometrics for Attention Estimation With Application to e-Learning}}}.
\newblock {\emph{\JournalTitle{IEEE Access}}} \textbf{\bibinfo{volume}{12}}, \bibinfo{pages}{111343--111359}, \url{10.1109/ACCESS.2024.3437291} (\bibinfo{year}{2024}).

\bibitem{altmann2014momentary}
\bibinfo{author}{Altmann, E.~M.}, \bibinfo{author}{Trafton, J.~G.} \& \bibinfo{author}{Hambrick, D.~Z.}
\newblock \bibinfo{journal}{\bibinfo{title}{{Momentary Interruptions can Derail the Train of Thought}}}.
\newblock {\emph{\JournalTitle{Journal of Experimental Psychology: General}}} \textbf{\bibinfo{volume}{143}}, \bibinfo{pages}{215}, \url{10.1037/a0030986} (\bibinfo{year}{2014}).

\bibitem{tanil2020mobile}
\bibinfo{author}{Tanil, C.~T.} \& \bibinfo{author}{Yong, M.~H.}
\newblock \bibinfo{journal}{\bibinfo{title}{{Mobile Phones: The Effect of its Presence on Learning and Memory}}}.
\newblock {\emph{\JournalTitle{PloS One}}} \textbf{\bibinfo{volume}{15}}, \bibinfo{pages}{e0219233}, \url{10.1371/journal.pone.0219233} (\bibinfo{year}{2020}).

\bibitem{mendoza2018effect}
\bibinfo{author}{Mendoza, J.~S.}, \bibinfo{author}{Pody, B.~C.}, \bibinfo{author}{Lee, S.}, \bibinfo{author}{Kim, M.} \& \bibinfo{author}{McDonough, I.~M.}
\newblock \bibinfo{journal}{\bibinfo{title}{{The Effect of Cellphones on Attention and Learning: The Influences of Time, Distraction, and Nomophobia}}}.
\newblock {\emph{\JournalTitle{Computers in Human Behavior}}} \textbf{\bibinfo{volume}{86}}, \bibinfo{pages}{52--60}, \url{10.1016/j.chb.2018.04.027} (\bibinfo{year}{2018}).

\bibitem{froese2012effects}
\bibinfo{author}{Froese, A.~D.} \emph{et~al.}
\newblock \bibinfo{journal}{\bibinfo{title}{{Effects of Classroom Cell Phone Use on Expected and Actual Learning}}}.
\newblock {\emph{\JournalTitle{College Student Journal}}} \textbf{\bibinfo{volume}{46}}, \bibinfo{pages}{323--332} (\bibinfo{year}{2012}).

\bibitem{cheever2014out}
\bibinfo{author}{Cheever, N.~A.}, \bibinfo{author}{Rosen, L.~D.}, \bibinfo{author}{Carrier, L.~M.} \& \bibinfo{author}{Chavez, A.}
\newblock \bibinfo{journal}{\bibinfo{title}{{Out of Sight is not Out of Mind: The Impact of Restricting Wireless Mobile Device Use on Anxiety Levels among Low, Moderate and High Users}}}.
\newblock {\emph{\JournalTitle{Computers in Human Behavior}}} \textbf{\bibinfo{volume}{37}}, \bibinfo{pages}{290--297}, \url{10.1016/j.chb.2014.05.002} (\bibinfo{year}{2014}).

\bibitem{hernandez2019edbb}
\bibinfo{author}{Hernandez-Ortega, J.}, \bibinfo{author}{Daza, R.}, \bibinfo{author}{Morales, A.}, \bibinfo{author}{Fierrez, J.} \& \bibinfo{author}{Ortega-Garcia, J.}
\newblock \bibinfo{title}{{{edBB: Biometrics and Behavior for Assessing Remote Education}}}.
\newblock In \emph{\bibinfo{booktitle}{Proc. AAAI Workshop on Artificial Intelligence for Education}} (\bibinfo{year}{2020}).

\bibitem{baro2018integration}
\bibinfo{author}{Bar{\'o}-Sol{\'e}, X.} \emph{et~al.}
\newblock \bibinfo{journal}{\bibinfo{title}{{Integration of an Adaptive Trust-based e-assessment System into Virtual Learning Environments—The TeSLA Project Experience}}}.
\newblock {\emph{\JournalTitle{Internet Technology Letters}}} \textbf{\bibinfo{volume}{1}}, \bibinfo{pages}{e56}, \url{10.1002/itl2.56} (\bibinfo{year}{2018}).

\bibitem{bhattacharjee2018enhancing}
\bibinfo{author}{Bhattacharjee, S.}, \bibinfo{author}{Ivanova, M.}, \bibinfo{author}{Rozeva, A.}, \bibinfo{author}{Durcheva, M.} \& \bibinfo{author}{Marcel, S.}
\newblock \bibinfo{title}{{{Enhancing Trust in eAssessment-the Tesla System Solution}}}.
\newblock In \emph{\bibinfo{booktitle}{Proc. on Technology Enhanced Assessment}} (\bibinfo{year}{2018}).

\bibitem{cobos2023self}
\bibinfo{author}{Cobos, R.}
\newblock \bibinfo{journal}{\bibinfo{title}{{Self-Regulated Learning and Active Feedback of MOOC Learners Supported by the Intervention Strategy of a Learning Analytics System}}}.
\newblock {\emph{\JournalTitle{Electronics}}} \textbf{\bibinfo{volume}{12}}, \bibinfo{pages}{3368}, \url{10.3390/electronics12153368} (\bibinfo{year}{2023}).

\bibitem{topali2024codesign}
\bibinfo{author}{Topali, P.}, \bibinfo{author}{Cobos, R.}, \bibinfo{author}{Agirre-Uribarren, U.}, \bibinfo{author}{Martínez-Monés, A.} \& \bibinfo{author}{Villagrá-Sobrino, S.}
\newblock \bibinfo{journal}{\bibinfo{title}{‘instructor in action’: Co-design and evaluation of human-centred la-informed feedback in moocs}}.
\newblock {\emph{\JournalTitle{Journal of Computer Assisted Learning}}} \url{10.1111/jcal.13057} (\bibinfo{year}{2024}).

\bibitem{daza2025smartevr}
\bibinfo{author}{Daza, R.}, \bibinfo{author}{Shengkai, L.}, \bibinfo{author}{Morales, A.}, \bibinfo{author}{Fierrez, J.} \& \bibinfo{author}{Nagao, K.}
\newblock \bibinfo{title}{{SMARTe-VR: Student Monitoring and Adaptive Response Technology for e-Learning in Virtual Reality}}.
\newblock In \emph{\bibinfo{booktitle}{Proc. AAAI Workshop on Artificial Intelligence for Education}} (\bibinfo{year}{2025}).

\bibitem{nagao2023virtual}
\bibinfo{author}{Nagao, K.}
\newblock \bibinfo{journal}{\bibinfo{title}{{Virtual Reality Campuses as New Educational Metaverses}}}.
\newblock {\emph{\JournalTitle{IEICE Transactions on Information and Systems}}} \textbf{\bibinfo{volume}{106}}, \bibinfo{pages}{93--100}, \url{10.1587/transinf.2022ETI0001} (\bibinfo{year}{2023}).

\bibitem{zhou2024stat}
\bibinfo{author}{Zhou, Y.}, \bibinfo{author}{Suzuki, K.} \& \bibinfo{author}{Kumano, S.}
\newblock \bibinfo{journal}{\bibinfo{title}{{State-Aware Deep Item Response Theory using Student Facial Features}}}.
\newblock {\emph{\JournalTitle{Frontiers in Artificial Intelligence}}} \textbf{\bibinfo{volume}{6}}, \bibinfo{pages}{1324279}, \url{10.3389/frai.2023.1324279} (\bibinfo{year}{2024}).

\bibitem{daza2024mebal2}
\bibinfo{author}{Daza, R.}, \bibinfo{author}{Morales, A.}, \bibinfo{author}{Fierrez, J.}, \bibinfo{author}{Tolosana, R.} \& \bibinfo{author}{Vera-Rodriguez, R.}
\newblock \bibinfo{journal}{\bibinfo{title}{{mEBAL2 Database and Benchmark: Image-based Multispectral Eyeblink Detection}}}.
\newblock {\emph{\JournalTitle{Pattern Recognition Letters}}} \textbf{\bibinfo{volume}{182}}, \bibinfo{pages}{83--89}, \url{10.1016/j.patrec.2024.04.011} (\bibinfo{year}{2024}).

\bibitem{becerra2023m2lads}
\bibinfo{author}{Becerra, {\'A}.} \emph{et~al.}
\newblock \bibinfo{title}{{M2LADS: A System for Generating MultiModal Learning Analytics Dashboards in Open Education}}.
\newblock In \emph{\bibinfo{booktitle}{Proc. Annual Computers, Software, and Applications Conference (COMPSAC) in the Workshop on Open Education Resources}}, \url{10.1109/COMPSAC57700.2023.00241} (\bibinfo{year}{2023}).

\bibitem{daza2021alebk}
\bibinfo{author}{Daza, R.} \emph{et~al.}
\newblock \bibinfo{title}{{ALEBk: Feasibility Study of Attention Level Estimation via Blink Detection applied to e-learning}}.
\newblock In \emph{\bibinfo{booktitle}{Proc. AAAI Workshop on Artificial Intelligence for Education}} (\bibinfo{year}{2022}).

\bibitem{rafiqi2015pupilware}
\bibinfo{author}{Rafiqi, S.} \emph{et~al.}
\newblock \bibinfo{title}{{{PupilWare: Towards Pervasive Cognitive Load Measurement using Commodity Devices}}}.
\newblock In \emph{\bibinfo{booktitle}{Proc. of the 8th ACM International Conference on Pervasive Technologies Related to Assistive Environments}}, \bibinfo{pages}{1--8}, \url{10.1145/2769493.2769506} (\bibinfo{year}{2015}).

\bibitem{intel2020realsense}
\bibinfo{author}{{Intel Corporation}}.
\newblock \bibinfo{title}{{Intel RealSense D400 Series Datasheet}}.
\newblock \bibinfo{howpublished}{\url{https://www.intelrealsense.com/wp-content/uploads/2020/06/Intel-RealSense-D400-Series-Datasheet-June-2020.pdf}} (\bibinfo{year}{2020}).
\newblock \bibinfo{note}{Accessed: 2025-06-10}.

\bibitem{hosseini2022multimodal}
\bibinfo{author}{Hosseini, S.} \emph{et~al.}
\newblock \bibinfo{journal}{\bibinfo{title}{{A Multimodal Sensor Dataset for Continuous Stress Detection of Nurses in a Hospital}}}.
\newblock {\emph{\JournalTitle{Scientific Data}}} \textbf{\bibinfo{volume}{9}}, \bibinfo{pages}{255}, \url{10.1038/s41597-022-01361-y} (\bibinfo{year}{2022}).

\bibitem{hernandez2020heart}
\bibinfo{author}{Hernandez-Ortega, J.}, \bibinfo{author}{Daza, R.}, \bibinfo{author}{Morales, A.}, \bibinfo{author}{Fierrez, J.} \& \bibinfo{author}{Tolosana, R.}
\newblock \bibinfo{title}{{Heart Rate Estimation from Face Videos for Student Assessment: Experiments on edBB}}.
\newblock In \emph{\bibinfo{booktitle}{Proc. of Annual Computers, Software, and Applications Conference}}, \bibinfo{pages}{172--177}, \url{10.1109/COMPSAC48688.2020.00031} (\bibinfo{year}{2020}).

\bibitem{cocskun2023physiological}
\bibinfo{author}{Co{\c{s}}kun, B.} \emph{et~al.}
\newblock \bibinfo{journal}{\bibinfo{title}{{A Physiological Signal Database of Children with Different Special Needs for Stress Recognition}}}.
\newblock {\emph{\JournalTitle{Scientific Data}}} \textbf{\bibinfo{volume}{10}}, \bibinfo{pages}{382}, \url{10.1038/s41597-023-02272-2} (\bibinfo{year}{2023}).

\bibitem{romero2023ai4fooddb}
\bibinfo{author}{Romero-Tapiador, S.} \emph{et~al.}
\newblock \bibinfo{journal}{\bibinfo{title}{{AI4FoodDB: a Database for Personalized e-Health Nutrition and Lifestyle through Wearable Devices and Artificial Intelligence}}}.
\newblock {\emph{\JournalTitle{Database}}} \textbf{\bibinfo{volume}{2023}}, \bibinfo{pages}{baad049}, \url{10.1093/database/baad049} (\bibinfo{year}{2023}).

\bibitem{acien2022becaptcha}
\bibinfo{author}{Acien, A.}, \bibinfo{author}{Morales, A.}, \bibinfo{author}{Fierrez, J.} \& \bibinfo{author}{Vera-Rodriguez, R.}
\newblock \bibinfo{journal}{\bibinfo{title}{{BeCAPTCHA-Mouse: Synthetic Mouse Trajectories and Improved Bot Detection}}}.
\newblock {\emph{\JournalTitle{Pattern Recognition}}} \textbf{\bibinfo{volume}{127}}, \bibinfo{pages}{108643}, \url{10.1016/j.patcog.2022.108643} (\bibinfo{year}{2022}).

\bibitem{kirschstein2009source}
\bibinfo{author}{Kirschstein, T.} \& \bibinfo{author}{K{\"o}hling, R.}
\newblock \bibinfo{journal}{\bibinfo{title}{{What is the Source of the EEG?}}}
\newblock {\emph{\JournalTitle{Clinical EEG and Neuroscience}}} \textbf{\bibinfo{volume}{40}}, \bibinfo{pages}{146--149}, \url{10.1177/155005940904000305} (\bibinfo{year}{2009}).

\bibitem{hall2020guyton}
\bibinfo{editor}{Hall, J.~E.} \& \bibinfo{editor}{Hall, M.~E.} (eds.) \emph{\bibinfo{title}{Guyton and Hall Textbook of Medical Physiology e-Book}} (\bibinfo{publisher}{Elsevier}, \bibinfo{year}{2020}).

\bibitem{chen2018effects}
\bibinfo{author}{Chen, C.-M.} \& \bibinfo{author}{Wang, J.-Y.}
\newblock \bibinfo{journal}{\bibinfo{title}{{Effects of Online Synchronous Instruction with an Attention Monitoring and Alarm Mechanism on Sustained Attention and Learning Performance}}}.
\newblock {\emph{\JournalTitle{Interactive Learning Environ}}} \textbf{\bibinfo{volume}{26}}, \bibinfo{pages}{427--443}, \url{10.1080/10494820.2017.1341938} (\bibinfo{year}{2018}).

\bibitem{li2011real}
\bibinfo{author}{Li, Y.} \emph{et~al.}
\newblock \bibinfo{title}{{A Real-Time EEG-based BCI System for Attention Recognition in Ubiquitous Environment}}.
\newblock In \emph{\bibinfo{booktitle}{Proc. Intl. Workshop on Ubiquitous Affective Awareness and Intelligent Interaction}}, \bibinfo{pages}{33--40}, \url{10.1145/2030092.2030099} (\bibinfo{year}{2011}).

\bibitem{daza2023matt}
\bibinfo{author}{Daza, R.} \emph{et~al.}
\newblock \bibinfo{title}{{{MATT: Multimodal Attention Level Estimation for e-learning Platforms}}}.
\newblock In \emph{\bibinfo{booktitle}{Proc. AAAI Workshop on Artificial Intelligence for Education}} (\bibinfo{year}{2023}).

\bibitem{lin2018mental}
\bibinfo{author}{Lin, F.-R.} \& \bibinfo{author}{Kao, C.-M.}
\newblock \bibinfo{journal}{\bibinfo{title}{{Mental Effort Detection using EEG Data in E-learning Contexts}}}.
\newblock {\emph{\JournalTitle{Computers \& Education}}} \textbf{\bibinfo{volume}{122}}, \bibinfo{pages}{63--79}, \url{10.1016/j.compedu.2018.03.020} (\bibinfo{year}{2018}).

\bibitem{chen2017assessing}
\bibinfo{author}{Chen, C.-M.}, \bibinfo{author}{Wang, J.-Y.} \& \bibinfo{author}{Yu, C.-M.}
\newblock \bibinfo{journal}{\bibinfo{title}{{Assessing the Attention Levels of Students by using a Novel Attention Aware System Based On Brainwave Signals}}}.
\newblock {\emph{\JournalTitle{British Journal of Educational Technology}}} \textbf{\bibinfo{volume}{48}}, \bibinfo{pages}{348--369}, \url{10.1111/bjet.12359} (\bibinfo{year}{2017}).

\bibitem{li2009towards}
\bibinfo{author}{Li, X.}, \bibinfo{author}{Hu, B.}, \bibinfo{author}{Zhu, T.}, \bibinfo{author}{Yan, J.} \& \bibinfo{author}{Zheng, F.}
\newblock \bibinfo{title}{{Towards Affective Learning with an EEG Feedback Approach}}.
\newblock In \emph{\bibinfo{booktitle}{Proc. of the First ACM International Workshop on Multimedia Technologies for Distance Learning}}, \bibinfo{pages}{33--38}, \url{10.1145/1631111.1631118} (\bibinfo{year}{2009}).

\bibitem{daza2020mebal}
\bibinfo{author}{Daza, R.}, \bibinfo{author}{Morales, A.}, \bibinfo{author}{Fierrez, J.} \& \bibinfo{author}{Tolosana, R.}
\newblock \bibinfo{title}{{{mEBAL: A Multimodal Database for Eye Blink Detection and Attention Level Estimation}}}.
\newblock In \emph{\bibinfo{booktitle}{Proc. Intl. Conf. on Multimodal Interaction}}, \bibinfo{pages}{32--36}, \url{10.1145/3395035.3425257} (\bibinfo{year}{2020}).

\bibitem{bagley1979effect}
\bibinfo{author}{Bagley, J.} \& \bibinfo{author}{Manelis, L.}
\newblock \bibinfo{journal}{\bibinfo{title}{{Effect of Awareness on an Indicator of Cognitive Load}}}.
\newblock {\emph{\JournalTitle{Perceptual and Motor Skills}}} \textbf{\bibinfo{volume}{49}}, \bibinfo{pages}{591--594}, \url{10.2466/pms.1979.49.2.591} (\bibinfo{year}{1979}).

\bibitem{sharma2016gaze}
\bibinfo{author}{Sharma, K.}, \bibinfo{author}{Alavi, H.~S.}, \bibinfo{author}{Jermann, P.} \& \bibinfo{author}{Dillenbourg, P.}
\newblock \bibinfo{title}{{A Gaze-based Learning Analytics Model: in-video Visual Feedback to Improve Learner's Attention in MOOCs}}.
\newblock In \emph{\bibinfo{booktitle}{Proc. Intl. Conference on Learning Analytics \& Knowledge}}, \bibinfo{pages}{417--421}, \url{10.1145/2883851.2883902} (\bibinfo{year}{2016}).

\bibitem{sharma2016visual}
\bibinfo{author}{Sharma, K.}, \bibinfo{author}{D'Angelo, S.}, \bibinfo{author}{Gergle, D.} \& \bibinfo{author}{Dillenbourg, P.}
\newblock \bibinfo{title}{{Visual Augmentation of Deictic Gestures in MOOC Videos}}.
\newblock In \emph{\bibinfo{booktitle}{Proc. Intl. Conference of the Learning Sciences}}, \bibinfo{pages}{202--209}, \url{10.22318/icls2016.28} (\bibinfo{year}{2016}).

\bibitem{andreu2016ealab}
\bibinfo{author}{Andreu-Perez, J.}, \bibinfo{author}{Solnais, C.} \& \bibinfo{author}{Sriskandarajah, K.}
\newblock \bibinfo{journal}{\bibinfo{title}{{EALab (Eye Activity Lab): a MATLAB Toolbox for Variable Extraction, Multivariate Analysis and Classification of Eye-Movement Data}}}.
\newblock {\emph{\JournalTitle{Neuroinformatics}}} \textbf{\bibinfo{volume}{14}}, \bibinfo{pages}{51--67}, \url{10.1007/s12021-015-9275-4} (\bibinfo{year}{2016}).

\bibitem{Navarro2024}
\bibinfo{author}{Navarro, M.} \emph{et~al.}
\newblock \bibinfo{title}{{VAAD: Visual Attention Analysis Dashboard Applied to e-Learning}}.
\newblock In \emph{\bibinfo{booktitle}{Proc. Intl. Symposium on Computers in Education (SIIE), IEEE}}, \bibinfo{pages}{1--6}, \url{10.1109/SIIE63180.2024.10604520} (\bibinfo{year}{2024}).

\bibitem{morales2016kboc}
\bibinfo{author}{Morales, A.} \emph{et~al.}
\newblock \bibinfo{title}{{KBOC: Keystroke Biometrics Ongoing Competition}}.
\newblock In \emph{\bibinfo{booktitle}{Proc. Intl. Conf. on Biometrics Theory, Applications and Systems}}, \bibinfo{pages}{1--6}, \url{10.1109/BTAS.2016.7791180} (\bibinfo{year}{2016}).

\bibitem{2016_IEEEAccess_KBOC_Aythami}
\bibinfo{author}{Morales, A.} \emph{et~al.}
\newblock \bibinfo{journal}{\bibinfo{title}{{Keystroke Biometrics Ongoing Competition}}}.
\newblock {\emph{\JournalTitle{IEEE Access}}} \textbf{\bibinfo{volume}{4}}, \bibinfo{pages}{7736--7746}, \url{10.1109/ACCESS.2016.2626718} (\bibinfo{year}{2016}).

\bibitem{BecerraM2LADSDEMO2025}
\bibinfo{author}{Becerra, A.}, \bibinfo{author}{Daza, R.}, \bibinfo{author}{Cobos, R.}, \bibinfo{author}{Morales, A.} \& \bibinfo{author}{Fierrez, J.}
\newblock \bibinfo{title}{{M2LADS Demo: A System for Generating Multimodal Learning Analytics Dashboards}}.
\newblock In \emph{\bibinfo{booktitle}{Proc. AAAI Workshop on Innovation and Responsibility in AI-Supported Education (iRAISE)}} (\bibinfo{year}{2025}).

\bibitem{bazarevsky2019blazeface}
\bibinfo{author}{{ Bazarevsky, Valentin and Kartynnik, Yury and Vakunov, Andrey and Raveendran, Karthik and Grundmann, Matthias}}.
\newblock \bibinfo{title}{{ Blazeface: Sub-millisecond Neural Face Detection on Mobile GPUs}}.
\newblock In \emph{\bibinfo{booktitle}{Proc. CVPR Workshop on Computer Vision for Augmented and Virtual Reality}} (\bibinfo{year}{2019}).

\bibitem{sandler2018mobilenetv2}
\bibinfo{author}{Sandler, M.}, \bibinfo{author}{Howard, A.}, \bibinfo{author}{Zhu, M.}, \bibinfo{author}{Zhmoginov, A.} \& \bibinfo{author}{Chen, L.-C.}
\newblock \bibinfo{title}{{{Mobilenetv2: Inverted Residuals and Linear Bottlenecks}}}.
\newblock In \emph{\bibinfo{booktitle}{Proc. of the IEEE Conference on Computer Vision and Pattern Recognition}}, \bibinfo{pages}{4510--4520}, \url{10.1109/CVPR.2018.00474} (\bibinfo{year}{2018}).

\bibitem{duan2019centernet}
\bibinfo{author}{Duan, K.} \emph{et~al.}
\newblock \bibinfo{title}{{Centernet: Keypoint Triplets for Object Detection}}.
\newblock In \emph{\bibinfo{booktitle}{Proc. of the IEEE/CVF international conference on computer vision}}, \bibinfo{pages}{6569--6578}, \url{10.1109/ICCV.2019.00667} (\bibinfo{year}{2019}).

\bibitem{zhou2020whenet}
\bibinfo{author}{{Zhou, Yijun and Gregson, James}}.
\newblock \bibinfo{title}{{Whenet: Real-time Fine-Grained Estimation for Wide Range Head Pose}}.
\newblock In \emph{\bibinfo{booktitle}{Proc. of the British Machine Vision Conference}} (\bibinfo{year}{2020}).

\bibitem{tan2019efficientnet}
\bibinfo{author}{{Tan, Mingxing and Le, Quoc}}.
\newblock \bibinfo{title}{{Efficientnet: Rethinking Model Scaling for Convolutional Neural Networks}}.
\newblock In \emph{\bibinfo{booktitle}{Proc. of the International Conference on Machine Learnin}}, \bibinfo{pages}{6105--6114} (\bibinfo{year}{2019}).

\bibitem{zhu2016face}
\bibinfo{author}{{Zhu, Xiangyu and Lei, Zhen and Liu, Xiaoming and Shi, Hailin and Li, Stan Z}}.
\newblock \bibinfo{title}{{Face Alignment across Large Poses: A 3d Solution}}.
\newblock In \emph{\bibinfo{booktitle}{Proc.of the IEEE Conference on Computer Vision and Pattern Recognition}}, \bibinfo{pages}{146--155}, \url{10.1109/CVPR.2016.23} (\bibinfo{year}{2016}).

\bibitem{joo2015panoptic}
\bibinfo{author}{{Joo, Hanbyul and Liu, Hao and Tan, Lei and Gui, Lin and Nabbe, Bart and Matthews, Iain and Kanade, Takeo and Nobuhara, Shohei and Sheikh, Yaser}}.
\newblock \bibinfo{title}{{Panoptic Studio: A Massively Multiview System for Social Motion Capture}}.
\newblock In \emph{\bibinfo{booktitle}{Proc. of the IEEE International Conference on Computer Vision}}, \bibinfo{pages}{3334--3342}, \url{10.1109/ICCV.2015.381} (\bibinfo{year}{2015}).

\bibitem{Daza2024IMPROVE}
\bibinfo{author}{Daza, R.}, \bibinfo{author}{Becerra, A.}, \bibinfo{author}{Cobos, R.}, \bibinfo{author}{Fierrez, J.} \& \bibinfo{author}{Morales, A.}
\newblock \bibinfo{title}{{IMPROVE dataset}}.
\newblock \bibinfo{howpublished}{\textit{Science Data Bank} \url{10.57760/sciencedb.14565}} (\bibinfo{year}{2024}).
\newblock \bibinfo{note}{Accessed: 2025-06-10}.

\bibitem{Becerra2024}
\bibinfo{author}{Becerra, A.} \emph{et~al.}
\newblock \bibinfo{title}{{Biometrics and Behavior Analysis for Detecting Distractions in e-Learning}}.
\newblock In \emph{\bibinfo{booktitle}{Proc. Intl. Symposium on Computers in Education (SIIE), IEEE}}, \bibinfo{pages}{1--6}, \url{10.1109/SIIE63180.2024.10604582} (\bibinfo{year}{2024}).

\bibitem{palatini1997heart}
\bibinfo{author}{Palatini, P.} \& \bibinfo{author}{Julius, S.}
\newblock \bibinfo{journal}{\bibinfo{title}{{Heart Rate and the Cardiovascular Risk}}}.
\newblock {\emph{\JournalTitle{Journal of Hypertensio}}} \textbf{\bibinfo{volume}{15}}, \bibinfo{pages}{3--17}, \url{10.1097/00004872-199715010-00001} (\bibinfo{year}{1997}).

\bibitem{hwang2019assessing}
\bibinfo{author}{Hwang, J.} \emph{et~al.}
\newblock \bibinfo{journal}{\bibinfo{title}{{Assessing Accuracy of Wrist-Worn Wearable Devices in Measurement of Paroxysmal Supraventricular Tachycardia Heart Rate}}}.
\newblock {\emph{\JournalTitle{Korean Circulation Journal}}} \textbf{\bibinfo{volume}{49}}, \bibinfo{pages}{437--445}, \url{10.4070/kcj.2018.0323} (\bibinfo{year}{2019}).

\bibitem{virtanen2020scipy}
\bibinfo{author}{Virtanen, P.} \emph{et~al.}
\newblock \bibinfo{journal}{\bibinfo{title}{{SciPy 1.0: Fundamental Algorithms for Scientific Computing in Python}}}.
\newblock {\emph{\JournalTitle{{Nature Methods}}}} \textbf{\bibinfo{volume}{17}}, \bibinfo{pages}{261--272}, \url{10.1038/s41592-020-0772-5} (\bibinfo{year}{2020}).

\bibitem{harris2020array}
\bibinfo{author}{Harris, C.~R.} \emph{et~al.}
\newblock \bibinfo{journal}{\bibinfo{title}{{Array Programming with NumPy}}}.
\newblock {\emph{\JournalTitle{Nature}}} \textbf{\bibinfo{volume}{585}}, \bibinfo{pages}{357--362}, \url{10.1038/s41586-020-2649-2} (\bibinfo{year}{2020}).

\bibitem{abadi2016tensorflow}
\bibinfo{author}{Abadi, M.} \emph{et~al.}
\newblock \bibinfo{title}{{$\{$TensorFlow$\}$: a System for $\{$Large-Scale$\}$ Machine Learning}}.
\newblock In \emph{\bibinfo{booktitle}{Proc. of the Symposium on Operating Systems Design and Implementation (OSDI 16)}}, \bibinfo{pages}{265--283} (\bibinfo{year}{2016}).

\bibitem{paszke2019pytorch}
\bibinfo{author}{Paszke, A.} \emph{et~al.}
\newblock \bibinfo{title}{{Pytorch: An Imperative Style, High-Performance Deep Learning Library}}.
\newblock In \emph{\bibinfo{booktitle}{Proc. of Advances in Neural Information Processing Systems}}, vol.~\bibinfo{volume}{32}, \bibinfo{pages}{8024--8035} (\bibinfo{year}{2019}).

\bibitem{king2009dlib}
\bibinfo{author}{King, D.~E.}
\newblock \bibinfo{journal}{\bibinfo{title}{{Dlib-ml: A Machine Learning Toolkit}}}.
\newblock {\emph{\JournalTitle{{The Journal of Machine Learning Research}}}} \textbf{\bibinfo{volume}{10}}, \bibinfo{pages}{1755--1758} (\bibinfo{year}{2009}).

\bibitem{lugaresi2019mediapipe}
\bibinfo{author}{Lugaresi, C.} \emph{et~al.}
\newblock \bibinfo{journal}{\bibinfo{title}{{MediaPipe: A Framework for Building Perception Pipelines}}}.
\newblock {\emph{\JournalTitle{arXiv preprint}}} \url{10.48550/arXiv.1906.08172} (\bibinfo{year}{2019}).

\bibitem{deng2020retinaface}
\bibinfo{author}{Deng, J.} \emph{et~al.}
\newblock \bibinfo{title}{{RetinaFace: Single-stage Dense Face Localisation in the Wild}}.
\newblock In \emph{\bibinfo{booktitle}{Proc. of the IEEE/CVF Conference on Computer Vision and Pattern Recognition (CVPR)}}, \bibinfo{pages}{5203--5212} (\bibinfo{year}{2020}).

\bibitem{berral2021realheponet}
\bibinfo{author}{Berral-Soler, R.}, \bibinfo{author}{Madrid-Cuevas, F.~J.}, \bibinfo{author}{Munoz-Salinas, R.} \& \bibinfo{author}{Mar{\'\i}n-Jim{\'e}nez, M.~J.}
\newblock \bibinfo{journal}{\bibinfo{title}{{RealHePoNet: a Robust Single-Stage ConvNet for Head Pose Estimation in the Wild}}}.
\newblock {\emph{\JournalTitle{Neural Computing and Applications}}} \textbf{\bibinfo{volume}{33}}, \bibinfo{pages}{7673--7689}, \url{10.1007/s00521-020-05511-4} (\bibinfo{year}{2021}).

\bibitem{peirce2007psychopy}
\bibinfo{author}{Peirce, J.~W.}
\newblock \bibinfo{journal}{\bibinfo{title}{{PsychoPy—psychophysics Software in Python}}}.
\newblock {\emph{\JournalTitle{Journal of Neuroscience Methods}}} \textbf{\bibinfo{volume}{162}}, \bibinfo{pages}{8--13}, \url{10.1016/j.jneumeth.2006.11.017} (\bibinfo{year}{2007}).

\end{thebibliography}
\end{document}